\documentclass[pss]{wiley2sp} 
\usepackage{amsmath}

\begin{document}

\title{ Spin-helical transport in normal and superconducting  topological insulators}

\titlerunning{Spin-helical transport in normal and superconducting topological insulators }

\author{%
  G. Tkachov \textsuperscript{\Ast} and E. M. Hankiewicz 
}

\authorrunning{ G. Tkachov and E. M. Hankiewicz }

\mail{e-mail
     \textsf{Grigory.Tkachov@physik.uni-wuerzburg.de} 
}

\institute{%
Institute for Theoretical Physics and Astrophysics, University of W\"urzburg, Am Hubland, 97074 W\"urzburg, Germany
}



\abstract{%
%
%
%
\abstcol{

In a topological insulator (TI) the character of electron transport varies from insulating in the interior of the material to metallic near its surface. 
Unlike, however, ordinary metals, conducting surface states in TIs are topologically protected and characterized by spin helicity 
whereby the direction of the electron spin is locked to the momentum direction. In this paper we review selected topics regarding recent theoretical and experimental work 
on electron transport and related phenomena in two-dimensional (2D) and three-dimensional (3D) TIs. 

}{%
The review provides a focused introductory discussion of the quantum spin Hall effect in HgTe quantum wells as well as transport properties of 3DTIs 
such as surface weak antilocalization, the half-integer quantum Hall effect, s + p-wave induced superconductivity, superconducting Klein tunneling, 
topological Andreev bound states and related Majorana midgap states. These properties of TIs are of practical interest, 
guiding the search for the routes towards topological spin electronics. 
}

}

%
%

\maketitle   

\section{Introduction.}
\label{Intro}

\begin{figure}[t]
\begin{center}
\includegraphics[width=75mm]{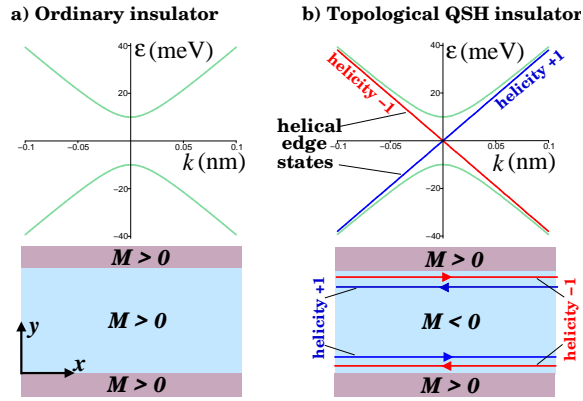}
\end{center}
\caption{
Schematic of band dispersion in 2D HgTe quantum wells: 
(a) ordinary band insulator with normal gap ${\cal M} > 0$ versus (b) 
topological quantum spin-Hall insulator (QSHI) with inverted gap ${\cal M}<0$. 
In the latter case, two crossing spectral branches correspond to a pair of 
edge states with opposite spin helicities $\tau = \pm 1$. We use Hamiltonian (\ref{H_HgTe_D})  
with ${\cal A}=380$ meV$\cdot$nm, $|{\cal M}|=10$ meV, ${\cal B}={\cal D}=0$ and boundary condition (\ref{BC}). 
}
\label{OI_TI}
\end{figure}

The situation when a material behaves as a metal in terms of its electric conductivity and, at the same time, as an insulator 
in terms of its band structure is extremely unconventional from the viewpoint of the standard classification of solids. 
That is why the recent discovery of a class of such materials - topological insulators (TIs) - has generated much interest 
(see e.g. reviews \cite{Hasan10,Qi11}). The dual properties of the TIs are especially well pronounced in two-dimensional (2D) systems 
which are also known as quantum spin-Hall insulators (QSHIs) \cite{Kane05,Bernevig06,Koenig07,Koenig08,Roth09}. 
In QSHIs, the metallic electric conduction is associated with propagating states that occur 
only near the sample edges, while the conduction in the interior is suppressed by a band gap like in ordinary band insulators.   
These edge states originate from intrinsic spin-orbit (SO) coupling and are profoundly different from those appearing in quantum Hall systems 
in a strong perpendicular magnetic field \cite{Halperin82,MacDonald84}. The key distinction lies in the role of the time reversal symmetry. 
In the QSHIs the SO coupling preserves the time-reversal symmetry, resulting in a pair of counter-propagating channels on the same edge as opposed 
to one-way directed (chiral) edge states in quantum Hall systems. Remarkably, the spin and momentum directions of the two QSHI edge channels are locked 
in the opposite ways so that these states are characterized by opposite spin helicities and orthogonal to each other. 
As a result, such helical edge states have a nodal band dispersion (see also Fig. \ref{OI_TI}) which is topologically protected against
any structural or sample imperfections that do not cause spin scattering. 

The QSHIs have been experimentally realized in HgTe/CdTe quantum wells with inverted band structure 
and strong intrinsic SO splitting coming from the atomically heavy mercury \cite{Koenig07,Koenig08,Roth09}. 
In these structures the helical edge states generate nonlocal transport effects in zero magnetic field 
that have no analogues in the conventional 2D semiconductors, allowing one to detect the helical edge states in appropriately designed Hall bar devices 
\cite{Roth09}. A variety of other properties of the QSHIs 
and, specifically, HgTe quantum wells have also got in the focus of recent theoretical 
 \cite{Yang08,Zhou08,GT09a,Yokoyama09,GT09b,Schmidt09,Akhmerov09,Groth09,Maciejko10a,GT10,Stroem10,Novik10,Guigou10,Adroguer10,Ostrovsky10,Meidan10,Rothe10,GT11_Res,Liu11_Dual,Schmidt11,GT11_Back,Chang11,Michetti11,Meidan11,Virtanen11,Wang11_QSH,DeMartino11,Weeks11,G_Liu11,Khaymovich11,Badiane11,Guigou11,Schaffer11,Timm11,Ruegg12,Virtanen12,Dora12,Budich12,Michetti12,Raichev12,GT12,JC_Chen12,Zou12,Beugeling12a,Goldman12,Kuzmenko12,Beugeling12b,Reinthaler12,Knez12,Shevtsov12,Schmidt12,Crepin12,Krueckl12,Ostrovsky12,Scharf12,Takagaki12_c}  
and experimental 
 \cite{Gusev10,Olshanetsky10,Bruene10,Buettner11,Ikonnikov11,Shuvaev11,Kvon12,Bruene12,Gusev12,Minkov12} research.

\begin{figure}[t]
\begin{center}
\includegraphics[width=35mm]{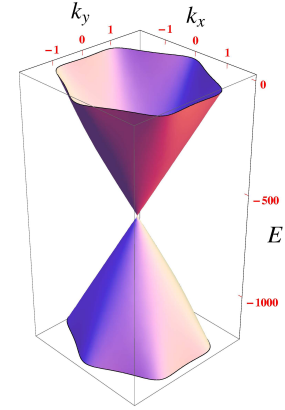}
\end{center}
\caption{
Energy bands (in meV) of a surface state in a 3DTI versus in-plane wave-numbers $k_x$ and $k_y$ (in nm$^{-1}$) 
from effective Hamiltonian (\ref{H_BiSe}). The Fermi level lies in the conduction band at $E=0$. 
We chose ${\cal A}=300$ meV$\cdot$nm, $W=50$ meV$\cdot$nm$^3$ and ${\cal D}=0$ meV$\cdot$nm$^2$ (adapted from \cite{GT11_WAL}).
}
\label{E_3D}
\end{figure}

The higher dimensional analogues of the QSHIs are the three-dimensional (3D) TIs \cite{Fu07a,Fu07b,Murakami07,Moore07,Hsieh08,Hsieh09,Zhang09}. 
In 3DTIs the topologically protected electronic states appear on the surface of a bulk material. 
These surface states have a nodal band dispersion in the form of a Dirac-like cone 
reflecting a continuum of momentum directions on the surface (see also Fig. \ref{E_3D}). 
The family of materials and heterostructures which can host surface states is pretty large. 
They were first predicted in inverted semiconductor contacts \cite{Volkov85,Pankratov87}.   
The coexistence of the metallic surface states with the bulk gapped band structure, i.e. the 3DTI phase, has been established theoretically 
for the semiconducting alloy Bi$_{1-x}$Sb$_x$ \cite{Fu07b}, strained 3D layers of $\alpha$-Sn and HgTe \cite{Fu07b}, 
the tetradymite semiconductors Bi$_2$Se$_3$, Bi$_2$Te$_3$, and Sb$_2$Te$_3$ \cite{Zhang09}, 
thallium-based ternary chalcogenides TlBiTe$_2$ and TlBiSe$_2$ \cite{Yan10,Lin10,Eremeev10a} as well as Pb-based layered chalcogenides \cite{Eremeev10b,Jin10}.  
Experimentally, topological surface states have been observed 
by means of angle-resolved photo-emission spectroscopy (ARPES) in Bi$_{1-x}$Sb$_x$ \cite{Hsieh08,Hsieh09}, Bi$_2$Se$_3$ \cite{Xia09}, 
Bi$_2$Te$_3$ \cite{Y_Chen09}, TlBiSe$_2$ \cite{Sato10,Kuroda10a,Y_Chen10}, 
TlBiTe$_2$ \cite{Y_Chen10}, Pb(Bi$_{1-x}$Sb$_x$)$_2$Te$_4$ \cite{Souma12} and PbBi$_2$Te$_4$ \cite{Kuroda12}.

Beside the ARPES and band structure calculations, there has been a growing number of experiments 
\cite{Checkelsky09,Taskin10,Butch10,Eto10,Analytis10,LaForge10,Sushkov10,J_Chen10,He11,Checkelsky11,Steinberg11,Kim11,J_Chen11,Taskin11,Bruene11,Bouvier11,Hancock11,D_Zhang11,Koren11,Sacepe11,Veldhorst12,Wang12_STI,Aguilar12,Cha12,Takagaki12,Takagaki12_b,Williams12,Shuvaev12} 
and theoretical studies 
\cite{Fu08,Qi08,Qi09,Akhmerov09_STI,Law09,Tanaka09,Essin09,Santos10,Linder10,Stanescu10,Mondal10b,Tse10a,Wu10,Sau10,Tse10b,Maciejko10b,Zazunov10,Culcer10,Bardarson10,Yazyev10,Chu11,Zhu11,Q_Li11,Dora11,Ioselevich11,Zyuzin11,Potter11,Labadidi11,Ito11,Culcer11a,Liu11_STI,GT11_TI,GT11_WAL,Nestoklon11,Lu11a,Lu11b,Burkov11a,Wang11,Golub11,Tse11,Culcer11b,Papic11,Tilahun11,Walter11,Vafek11,Fu11,Beenakker11,Tanaka12,Tserkovnyak12,Culcer12,Abanin12,Pesin12,Beri12,Tewari12,Bergman12,Adroguer12,Garate12,Tudorovskiy12,Olund12,Tahir12,She12,Bardarson12,Tian12}   
devoted to helical transport in 3DTI materials. 
Like in the QSHIs, the surface charge carriers in 3DTIs are characterized by a well-defined spin helicity, 
i.e. the locking of the spin and momentum directions. 
This has several implications for electron transport which will be the subject of the present review.  
First, the helicity conservation along closed electron trajectories prevents ordinary electron localization on the surface. Instead, experiments 
\cite{Checkelsky09,J_Chen10,Checkelsky11,He11,Bouvier11,Steinberg11,Kim11,J_Chen11,Cha12,Takagaki12,Takagaki12_b} 
and theoretical calculations \cite{GT11_WAL,Nestoklon11,Lu11a,Lu11b,Adroguer12,Garate12} have indicated that the surface quantum transport in the 3DTIs 
should exhibit weak antilocalization with a positive magnetoresistance. In thin TI films with surface and bulk states, both weak antilocaliation and localization 
regimes have been predicted \cite{Lu11b,Garate12}.
Second, in a strong perpendicular magnetic field the unconventional half-integer quantum Hall plateaus are expected for a single spin-helical surface 
\cite{Fu07b,Qi08,Tse10a,Tse10b,Tse11}, as has been found in graphene \cite{Neto09} and 
in zero-gap HgTe/CdTe quantum wells \cite{Buettner11}. Experimentally, unusual odd and even quantum Hall plateaus 
have been observed in epitaxially strained 3DTI HgTe \cite{Bruene11}, which may
indicate the contribution of two spin-helical surfaces, one at the top and one at bottom of the sample.
Beside the magnetotransport, the quantum Hall dynamics in 3DTIs manifests itself in the Faraday effect at THz frequencies \cite{Shuvaev12}.
Another interesting implication of the spin helicity is the possibility of 
unconventional surface superconductivity. It is expected to occur in proximity of a singlet s-wave superconductor (e.g. Nb or Al) which induces 
both singlet s-wave and triplet p-wave correlations, as a result of the broken spin-rotation symmetry \cite{Fu08,Santos10,Stanescu10,Potter11,Labadidi11,Tanaka12}.
As a signature of the p-wave correlations, topological Majorana midgap states have been predicted in superconductor (S)/3DTI junctions 
(see e.g. \cite{Fu08,Qi11,Beenakker11,Tanaka12}). The theoretical search for unconventional superconductivity in 3DTIs goes in parallel with 
experimental progress in fabrication and characterization of such hybrid superconducting systems 
\cite{D_Zhang11,Koren11,Sacepe11,Veldhorst12,Wang12_STI,Williams12}.  

\begin{figure}[t]
\begin{center}
\includegraphics[width=75mm]{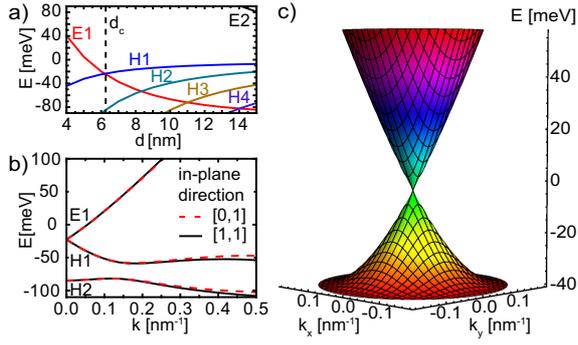}
\end{center}
\caption{
Band structure of a HgTe/Hg$_{0.3}$Cd$_{0.7}$Te quantum well:
(a) electron $E1, E2,...$ and heavy-hole $H1, H2, ...$ subband energies versus well thickness $d$, 
(b) in plane dispersion at the critical thickness $d_c\approx 6.3$ nm, and
(c) a 3D plot of the Dirac-like low-energy spectrum for $d=d_c$ near the $\Gamma$ point of the Brillouin zone (adapted from \cite{Buettner11}).  
}
\label{HgTe}
\end{figure}

As transport in TIs is a multifaceted problem, this review aims to discuss it from different angles, starting from   
the QSH effect and related phenomena in 2D HgTe quantum wells and continuing to the surface weak antilocalization, half-integer quantum Hall effect and 
s + p-wave induced superconductivity in 3DTIs. Regarding the superconducting properties, we introduce a generalized Fu-Kane model correctly accounting 
for the energy-dependence of the proximity effect in 3DTIs. We then analyze the superconducting Klein tunneling, topological Andreev bound states and 
related Majorana midgap states as well as the fractional AC Josephson effect \cite{Kitaev01,Kwon04} in such proximity S/TI/S junctions.

\section{ Topological insulators in two dimensions: Quantum spin Hall insulators.}
\label{2DTI}

\subsection{Effective Hamiltonian for HgTe QWs.}
\label{HgT_Model}
We will discuss the QSHIs in the context of their experimental realization in HgTe/CdTe quantum wells (QWs) \cite{Koenig07,Koenig08,Roth09}.   
HgTe is a zinc-blende-type semiconductor. Due to large relativistic corrections, it has an inverted band structure 
where a metallic s-type electron band (usually acting as the conduction band) has a lower energy than p-type hole bands \cite{Chu08}.  
Consequently, in HgTe QWs with large thicknesses $d$ the energy subbands are also inverted (see Fig. \ref{HgTe}a). 
In Fig. \ref{HgTe}, subbands $H1, H2, ...$ originate from the heavy-hole band, whereas the electron-like subbands are denoted by $E1, E2,...$. 
With the decreasing QW thickness $d$, the energies of the $E$ subbands increase as a result of quantum confinement, 
whereas those of the $H$ subbands decrease (see Fig. \ref{HgTe}a). 
This results in the normal band sequence in narrow QWs. 
The different $d$-dependences of $E1$ and $H1$ subbands imply a critical thickness, $d_c$, 
at which the band gap is closed. For $d \approx d_c$ and near the $\Gamma$ point, an effective four-band model 
involving double (Kramers) degenerate $E1$ and $H1$ subbands can be derived from the eight-band Kane model~\cite{Bernevig06,Rothe10}.  
Introducing basis states $|E1 +\rangle$, $|H1 +\rangle$, $|E1 -\rangle$ and $|H1 -\rangle$ (where $\pm$ denotes Kramers partners), 
one can write the effective four-band Hamiltonian for the system as follows~\cite{Bernevig06,Rothe10}:
\begin{equation}
H_{HgTe}=
\Biggl[
  \begin{array}{cc}
    h({\bf k}) & 0 \\
    0 & h^{\ast}({\bf -k}) \\
  \end{array}
\Biggr],
\label{H_HgTe}
\end{equation}
\begin{equation}
h({\bf k}) = {\cal A} (\sigma_x k_x - \sigma_y k_y) + {\cal M}_{\bf k}\sigma_z + {\cal D}{\bf k}^2 \sigma_0,
\label{h_HgTe}
\end{equation}
\begin{equation}
{\cal M}_{\bf k} = {\cal M} + {\cal B}{\bf k}^2.
\label{M}
\end{equation}
The two diagonal blocks of $H_{HgTe}$ (\ref{H_HgTe}) describe pairs of states related to each other by time reversal symmetry (Kramers partners). 
Each of the blocks has a matrix $2 \times 2$ structure with Pauli matrices $\sigma_{x,y,z}$ and 
unit matrix $\sigma_0$ representing the two lowest-energy subbands $E1$ and $H1$. 
The linear terms in Eq. (\ref{h_HgTe}) (proportional to constant ${\cal A}$ and in-plane wave-vectors $k_{x,y}$) 
describe the $E1$-$H1$ hybridization, while ${\cal M}_{\bf k}$ yields the band gap ${\cal M}$ 
at the $\Gamma$ (${\bf k}=0$) point of the Brillouin zone. The positive quadratic terms ${\cal B}{\bf\hat k}^2$ and ${\cal D}{\bf\hat k}^2$ 
take into account the details of the band curvature in HgTe QWs~\cite{Bernevig06}. 
The Hamiltonian (\ref{H_HgTe}) can be extended to include the spin-orbit coupling between the Kramers partners~\cite{Rothe10,Koenig08}. 

Using unitary transformation $H \to U H U^\dagger$ with 
$
U = \bigl(
  \begin{smallmatrix}
    0 &  \sigma_z \\
    -i\sigma_y & 0 
  \end{smallmatrix}
\bigr),
$
we can cast the Hamiltonian (\ref{H_HgTe}) into a Dirac-like form  
\begin{equation}
H = \tau_z \mbox{\boldmath$\sigma$} \cdot ( {\cal A} {\bf k} + {\cal M}_{\bf k} {\bf z} ) +  {\cal D}{\bf k}^2 \tau_0\sigma_0, 
\label{H_HgTe_D}
\end{equation}
where Pauli matrix $\tau_z$ and unit matrix $\tau_0$ act on the Kramers partners. 
We note that despite the effective mass term ${\cal M}_{\bf k}\tau_z\sigma_z$ the Hamiltonian (\ref{H_HgTe_D}) is invariant under time reversal, 
i.e. ${\cal T}^\dagger H {\cal T} = H$, where ${\cal T} = i\tau_y \, \sigma_x \, {\cal C}$ is the time-reversal operator, 
with ${\cal C}$ denoting complex conjugation.

\subsection{Simple analytic model of the QSH insulator.}
\label{QSHI}
The QSHI state is realized when a QW with an inverted gap ${\cal M} < 0$ 
is sided by ordinary band insulators with ${\cal M} > 0$ (see also Fig. \ref{OI_TI}b). 
The system boundaries play a role of topological defects - "mass" domain walls - 
that bind electronic states near the edge so that they decay on both sides of the boundary. 
In this subsection we discuss specific properties of such edge states: 
\begin{itemize}
\item 
the edge-state spectrum is gapless and merges into the bulk spectrum above the band gap;  

\item 
the edge-states are the orthogonal eigenstates of the helicity operator $\Sigma = \tau_z \mbox{\boldmath$\sigma$}\cdot \hat{\bf k}$, 
where $\hat{\bf k}$ is the unit vector in the direction of the edge-state momentum. For this reason, the QSH edge states are called helical;  

\item 
local static perturbation $V$ preserving time-reversal symmetry does not couple the QSH edge states.   
\end{itemize}
In order to illustrate these properties we will make two simplifications. 
First, we will omit all the terms $\propto {\bf k}^2$ in Hamiltonian (\ref{H_HgTe_D}). 
This is justified since the edge states occur in the vicinity of the $\Gamma ({\bf k}=0)$ point. 
Hamiltonian (\ref{H_HgTe_D}) takes, then, the form  
$
H = \tau_z \mbox{\boldmath$\sigma$} \cdot ( {\cal A} {\bf k} + {\cal M} {\bf z} ), 
$
which in position representation corresponds to the following equation for the four-component wave function $\Psi({\bf r})$:
\begin{equation}
 [\epsilon \sigma_0 - \tau_z \mbox{\boldmath$\sigma$} \cdot ( -i{\cal A} \nabla + {\cal M} {\bf z} )]\Psi({\bf r}) = 0.
\label{Eq_D}
\end{equation}
Second, we will assume that our system is confined by a normal band insulator with the infinite mass, $M \to +\infty$. 
It is known \cite{Berry87} that such infinite mass confinement can be modeled by an effective local boundary condition which in our geometry (e.g. at $y=0$) reads     
\begin{equation} 
\Psi(x,y=0) = \tau_0\sigma_x\, \Psi(x,y=0).
\label{BC}
\end{equation}
This boundary condition is specific to Dirac fermions with linear spectrum. It ensures vanishing of the normal component of the particle current without putting 
$\Psi(x,y)$ to zero at the boundary. The use of the infinite mass confinement (\ref{BC}) is complementary to  
the tight-binding calculations (see e.g. Refs.\cite{Koenig08,Maciejko10a}) and other continuum models of the helical edge states, 
which include the quadratic (${\bf k}^2$) terms and employ the hard wall boundary conditions (see e.g. Refs. \cite{Zhou08,Yokoyama09}).

We seek solutions to Eq. (\ref{Eq_D}) in form of the two eigenstates, $\Psi_{ k,\pm }({\bf r})$, of diagonal matrix $\tau_z\sigma_0$ 
propagating along the edge (in the $x$-direction) and decaying exponentially away from it (in the $y$-direction): 
\begin{eqnarray}
&&
\Psi_{k,+}({\bf r}) = 
\left(
  \begin{array}{c}
     1 \\
     0 \\
  \end{array}
\right)
\otimes   
\left(
  \begin{array}{c}
     \Psi_{1k+}\\
     \Psi_{2k+} \\
  \end{array}
\right)
{\rm e}^{ ikx - y/\lambda},
\label{Ansatz+}\\
%
&&
\Psi_{k,-}({\bf r}) = 
\left(
  \begin{array}{c}
     0 \\
     1 \\
  \end{array}
\right)
\otimes   
\left(
  \begin{array}{c}
     \Psi_{1k-}\\
     \Psi_{2k-} \\
  \end{array}
\right)
{\rm e}^{ikx - y/\lambda},
\label{Ansatz-}
\end{eqnarray}
with a real positive decay length $\lambda > 0$. 
The symbol $\otimes$ denotes a tensor product of an eigenstate of $\tau_z$ (first column) 
and the wave function in $\sigma$ space (second column). 
The conditions for the nontrivial solutions for the coefficients $\Psi_{1k\pm}$ and $\Psi_{2k\pm}$ 
follow from Eqs. (\ref{Eq_D}) and (\ref{BC}), 
yielding two equations for $\lambda$ and $\epsilon$:
\begin{eqnarray}
&& \lambda^{-2} - M^2/{\cal A}^2  = k^2 - \epsilon^2/{\cal A}^2,
\label{Eq1}\\
%
&& \lambda^{-1} + M/{\cal A} = k - \epsilon/{\cal A}\tau, \quad \tau=\pm 1.
\label{Eq2}
\end{eqnarray}
We notice that the left-hand-side of Eq. (\ref{Eq2}) does not contain index $\tau$, whereas the right-hand-side does. 
This can only be true if both sides of Eq. (\ref{Eq2}) [and those Eq. (\ref{Eq1})] vanish independently, 
which yields a solution with a gapless linear dispersion and a real decay length:
\begin{equation}
\epsilon_{k \tau} = {\cal A}  k \tau, \qquad \lambda = - {\cal A}/{\cal M}, 
\qquad {\cal M} <0.
 \label{E_edge}
\end{equation}
Since $\lambda$ must be positive, the edge states exists only in a system with the inverted negative gap, disappearing when ${\cal M}$ turns positive. 
Their propagation velocity $v={\cal A}/\hbar$ coincides with that of the bulk states above the gap (see also Fig. \ref{OI_TI}).
In a narrow QSHI the overlap of the edge states from the opposite sides results in a gapped edge-state dispersion \cite{Zhou08}.  

The edge-state wave functions normalized to half-space $0 \leq y < \infty$ are given by
\begin{eqnarray}
&&
\Psi_{k,+}({\bf r}) = 
\left(
  \begin{array}{c}
     1 \\
     0 \\
  \end{array}
\right)
\otimes   
\left(
  \begin{array}{c}
     1\\
     1\\
  \end{array}
\right)
\sqrt{ \frac{|{\cal M}|}{ {\cal A} } }\,
{\rm e}^{ ikx -  |{\cal M}|y/\cal A},
\label{Edge+}\\
%
&&
\Psi_{k,-}({\bf r}) = 
\left(
  \begin{array}{c}
     0 \\
     1 \\
  \end{array}
\right)
\otimes   
\left(
  \begin{array}{c}
     1\\
     1\\
  \end{array}
\right)
\sqrt{ \frac{|{\cal M}|}{ {\cal A} } }\,
{\rm e}^{ ikx -  |{\cal M}|y/\cal A}.
\label{Edge-}
\end{eqnarray}
The key feature of the edge states (\ref{Edge+}) and (\ref{Edge-}) 
is that they are orthogonal eigenstates of the helicity operator $\Sigma = \tau_z \sigma_x$: 
\begin{equation}
\Sigma \, \Psi_{k,\tau}({\bf r}) = \tau \, \Psi_{k,\tau}({\bf r}), \qquad  \tau=\pm 1.
\label{Helicity}
\end{equation}
The helicity $\Sigma$ is defined as the projection of vector $\mbox{\boldmath$\Sigma$} = \tau_z \mbox{\boldmath$\sigma$}$ 
on the direction of the edge-state momentum $\hat{\bf k} \| x$. Since the matrix structure of $\Sigma$ derives from the SO-split energy bands, 
it is also called the spin helicity. Equation (\ref{Helicity}) is a manifestation of the time-reversal symmetry and the fact that the QSH state is generally 
characterized by a $Z_2$ topological invariant \cite{Kane05}. It is easy to see that one helical channel can be obtained from the other 
by simply applying the time-reversal operator 
${\cal T} = i\tau_y \, \sigma_x \, {\cal C}$: 
\begin{equation}
 \Psi_{k,+} = {\cal T} \, \Psi_{-k,-}, \qquad \Psi_{-k,-} = -{\cal T} \, \Psi_{k,+}.
\label{TR_pm}
\end{equation}

The helicity [see Eq. (\ref{Helicity})] protects the edge states from local perturbations that does not break the time-reversal symmetry. 
Concretely, let us consider a perturbation $V$ which is (up to a phase) invariant under time reversal:   
\begin{equation}
{\cal T}^\dagger \, V \, {\cal T} = V^*.
 \label{V}
\end{equation}
Using Eqs. (\ref{TR_pm}) and (\ref{V}) we can transform 
the matrix element $\langle -k,-| V |k,+\rangle$ as follows
\begin{equation}
\langle -k,-| V |k,+\rangle = - \langle k,+| {\cal T}^\dagger \, V \, {\cal T} |-k,-\rangle =  
\label{V_pm1}
\end{equation}
\begin{equation}
= - \langle k,+| V^* |-k,-\rangle = - \langle -k,-| V^\dagger |k,+\rangle,
 \label{V_pm2}
\end{equation}
i.e. for a hermitian $V=V^\dagger$ its matrix element is zero:
\begin{equation}
\langle -k,-| V |k,+\rangle = 0.
\label{V_pm3}
\end{equation}
In particular, spin-independent potential disorder cannot cause scattering between the helical edge states and their localization. 
The physics of localization in the QSHIs has been studied by using both field-theoretical \cite{Stroem10,Ostrovsky10} and numerical 
(see e.g. \cite{Maciejko10a,Wang11_QSH,Weeks11,Zou12,JC_Chen12}) techniques. In particular, Ref. \cite{Maciejko10a} 
has found that the edge backscattering and magnetoresistance can occur as a combined effect of bulk-inversion asymmetry, 
sufficiently strong potential disorder and an external magnetic field. 

Another question regarding the role of the time-reversal symmetry is what happens to the QSHIs in a strong quantizing magnetic field? 
In this case the helical edge states appear within the gap between the highest hole ($-|{\cal M}|, \tau = +1$) and 
lowest particle ($|{\cal M}|, \tau=-1$) Landau levels \cite{GT10}. The edge-state dispersion is still nodal, but nonlinear, i.e. 
the two crossing spectral branches have now different group velocities. The latter circumstance makes the helical edge states prone to 
get coupled by spin-flip scattering, which also generates the edge magnetoresistance \cite{GT10,GT12}.

We should emphasize that Eqs. (\ref{V_pm1}) - (\ref{V_pm3}) do not generally hold for interacting helical channels where the electron backscattering 
may occur provided that the axial spin rotation symmetry of Hamiltonian (\ref{H_HgTe}) is broken, e.g. by the Bychkov-Rashba spin-orbit coupling 
\cite{Stroem10,Budich12,Schmidt12}. The spatially random  Bychkov-Rashba coupling has been predicted to cause localization of the edge states 
in the presence of weakly screened electron-electron interaction \cite{Stroem10}. Ref. \cite{Schmidt12} 
has shown that weak electron-electron interaction 
in the absence of the axial spin rotation symmetry allows for inelastic backscattering of a single electron, 
accompanied by forward scattering of another. 
The inelastic phonon-induced backscattering in the helical liquid has been considered in Ref. \cite{Budich12}.

\begin{figure}[t]
\begin{center}
\includegraphics[width=55mm]{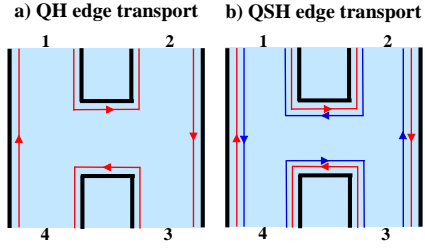}
\end{center}
\caption{
Schematic of edge transport in four-terminal device: 
(a) chiral edge states in a quantum Hall (QH) system versus 
(b) helical edge states in the quantum spin Hall (QSH) insulator.
}
\label{QH_QSH}
\end{figure}

\subsection{Helical versus chiral edge transport. Nonlocal detection of the QSHI state.}
\label{Nonlocal}

The helical edge transport in a QSHI differs markedly from the chiral edge transport in QH systems. 
In order to illustrate this we follow Ref. \cite{Roth09} and consider a four-terminal device shown in Fig. \ref{QH_QSH}. 
Using the ballistic Landauer-B\"utikker approach, we express the current, $I_i$, injected through contact $i$ in terms of voltages $V_j$ 
induced on all contacts as  
\begin{equation}
I_i = \frac{e^2}{h} \sum_{j=1}^N (T_{ji} V_i - T_{ij} V_j),
\label{I}
\end{equation}
where $T_{ji}$ is the transmission probability from contact i to contact j. 
For a chiral QH edge channel, $T_{ji}$ connects the neighboring contacts only 
in one propagation direction (see also Fig. \ref{QH_QSH}a), such that     
\begin{equation}
T(QH)_{i+1,i} = 1, \qquad i=1,...,N,  
\label{T_QH}
\end{equation}
where $N$ is the number of the terminals (e.g. $N=4$ in Fig. \ref{QH_QSH}), 
with the convention that $T_{N+1,N} = T_{1,N}$ describes the transmission from terminal $N$ to terminal 1. 
Assuming, for concreteness, that the current flows from terminal 1 to terminal 4, while leads 2 and 3 are used as voltage probes, 
we have  
\begin{equation}
I_1 \equiv I_{14} = \frac{e^2}{h} (V_1 - V_4),\,\, V_2 = V_1, \,\, V_3 =V_2, \,\, I_4 = - I_1.
\label{QH_1234}
\end{equation}
This yields a finite two-terminal resistance $R_{14,14} =\frac{V_1 - V_4}{ I_{14} }= h/e^2$ and 
zero four-terminal (nonlocal) resistances $R_{14,12} =\frac{V_1 - V_2}{ I_{14} }=0$, $R_{14,23} =\frac{ V_2 - V_3 }{ I_{14} }=0$ 
and $R_{14,13} =\frac{ V_1 - V_3 }{ I_{14} }=0$. 

In contrast, in a QSHI the helical edge channels connect the neighboring contacts in both propagation directions 
(see also Fig. \ref{QH_QSH}b), such that   
\begin{equation}
T(QSH)_{i+1,i} = T(QSH)_{i, i+1} = 1, \qquad i=1,...,N,  
\label{T_QSH}
\end{equation}
with the conventions $T_{N+1,N} = T_{1,N}$ and $T_{N, N+1} = T_{N, 1}$. 
Consequently, for a current flowing from 1 to 4, we find 
\begin{eqnarray}
&&
I_1 \equiv I_{14} = \frac{e^2}{h} (2V_1 - V_4 - V_2 ),\qquad I_4 = - I_1,
\label{QSH_14}\\
&&
2V_2 - V_1 - V_3 =0, \qquad 2V_3 - V_2 - V_4 =0,
\label{QSH_23}
\end{eqnarray}
which yields the two-terminal resistance \cite{Roth09}
\begin{equation}
R_{14,14} =\frac{V_1 - V_4}{ I_{14} } = \frac{3}{4} \frac{h}{e^2},
\label{R14_14}
\end{equation}
and the four-terminal resistances \cite{Roth09}
\begin{equation}
R_{14,12} =\frac{V_1 - V_2}{ I_{14} }=R_{14,23} =\frac{ V_2 - V_3 }{ I_{14} } = \frac{1}{4} \frac{h}{e^2},
\label{R14_12_23}
\end{equation}
\begin{equation}
R_{14,13} =\frac{V_1 - V_3}{ I_{14} }=\frac{1}{2} \frac{h}{e^2}.
\label{R14_13}
\end{equation}
The nonzero non-local resistances (\ref{R14_12_23}) and (\ref{R14_13}) 
are unique to the QSHI state, allowing its unambiguous experimental detection \cite{Roth09}.  
It should be emphasized that the universality of the non-local resistances (\ref{R14_12_23}) and (\ref{R14_13}) is just 
the consequence of the time-reversal symmetry and, therefore, is expected also for other proposed realizations of the QSHIs, e.g. 
in inverted InAs/GaSb quantum wells \cite{Knez12}. 
Equations (\ref{R14_14}) -- (\ref{R14_13}) for the quantized resistances 
are valid in the zero-temperature limit when the inelastic backscattering processes are negligible \cite{Kane05}.  

\subsection{Helical carriers and weak antilocalization in n-type HgTe quantum wells.}
\label{n-type}

\begin{figure}[t]
\begin{center}
\includegraphics[width=70mm]{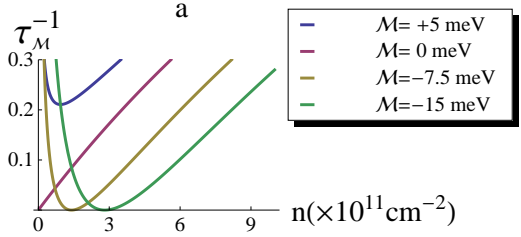}
\end{center}
\caption{
Backscattering rate $\tau^{-1}_{\cal M}$ [in units of elastic scatering rate $\tau^{-1}$, see Eq.~(\ref{Gamma})] versus 
carrier density $n$;  ${\cal A}=380$ meV$\cdot$nm and ${\cal B}=850$ meV$\cdot$nm$^2$ (adapted from \cite{GT11_WAL}). 
}
\label{Rate}
\end{figure}

The absence of the edge localization in the QSHI regime (i.e. when the Fermi level lies in the band gap) is 
closely related to the weak antilocalization (WAL) effect observed when the Fermi level is pushed above the gap into the bulk conduction band. 
The latter case corresponds to n-type HgTe quantum wells in which charge carriers behave as a 2D helical metal. 
Assuming that it is described by the same Hamiltonian (\ref{H_HgTe_D}), we can calculate the disorder-induced quantum-interference 
correction $\delta\sigma$ to the classical Drude conductivity (for more details see Sec. \ref{3DTIs}). 
In the leading logarithmic order  $\delta\sigma$ is given by \cite{GT11_WAL}  
\begin{equation}
\delta\sigma_{xx}(n,{\cal M}) \approx  \frac{2e^2} {2 \pi h} \ln\frac{\tau^{-1}}{ \tau^{-1}_{\cal M} + \tau^{-1}_\varphi },
\label{dS_HgTe}
\end{equation}
\begin{equation} 
\tau^{-1}_{\cal M} = \frac{2}{\tau} \left( \frac{\gamma_B - \pi}{\pi} \right)^2, 
\label{Gamma}
\end{equation}
\begin{equation}
\gamma_B =\pi \left(1 + \frac{ {\cal M} + {\cal B}k^2_F   }{ \sqrt{ {\cal A}^2k^2_F + ( {\cal M} + {\cal B}k^2_F )^2 } } \right),
k_F=\sqrt{2\pi n}.
\label{Berry}
\end{equation}
where $\tau$ is the elastic scattering time and $\tau_\varphi$ is the dephasing time. 
The third new time-scale $\tau_{\cal M}$ appears due to the lack of topological protection of the gapped spectrum:
its geometrical Berry phase $\gamma_B$ (\ref{Berry}) no longer coincides with the universal value $\pi$ characteristic of a massless Dirac cone \cite{Ando05}. 
The presence of the gap ${\cal M} + {\cal B}k^2_F$ enables scattering between opposite-momenta states ${\bf k}_F$ and $-{\bf k}_F$ (backscattering) on the 2D Fermi surface 
(see also Ref. \cite{GT11_Back}). Thus, $\tau^{-1}_{\cal M}$ is the rate of such backscattering. It is an interesting band-structure effect (see also Fig. \ref{Rate}): 
for the normal band structure with ${\cal M}> 0$ the rate $\tau^{-1}_{\cal M}$ is finite at any carrier density $n$,
whereas for the inverted band structure with ${\cal M}< 0$ the backscattering rate $\tau^{-1}_{\cal M}$ is strongly nonmonotonic with a zero 
at a specific carrier density $n = -  {\cal M}/ 2\pi {\cal B}$. 

The role of the geometrical phases in transport quantum-interference effects in HgTe quantum wells 
has recently been studied numerically in Ref. \cite{Krueckl12}.
In Ref. \cite{Ostrovsky12} analytical results for the WAL in HgTe quantum wells have been obtained with the account of 
both bulk-inversion asymmetry and Rashba SO coupling. Experimentally, the WAL in HgTe has been observed 
in 2D quantum wells \cite{Olshanetsky10,Minkov12} and in strained 3D layers \cite{Bouvier11}. 
Also, experiments have revealed cyclotron resonance phenomena for the helical carriers 
in HgTe quantum wells \cite{Ikonnikov11,Kvon12}.

\section{ Spin-helical transport in 3D topological insulators. }
\label{3DTIs}

The 1D edge states of HgTe/CdTe quantum wells discussed above have higher-dimensional analogues, such as 
gapless states on a 2D interface between two bulk semiconductors with normal and inverted gaps \cite{Volkov85,Pankratov87}. 
Generally, a 3D system with gapless surface states occuring inside the bulk band gap 
realizes a topogically nontrivial insulating state, the 3DTI \cite{Fu07a,Fu07b,Murakami07,Moore07,Zhang09,Hasan10,Qi11}. 
Since the gapless states in 3DTIs appear only on the surface, they are exempt from the fermion doubling theorem \cite{Nielsen81}.  
For this reason, the topological surface state consists of an odd number of Weyl-like fermions, 
each described by a two-component spinor wave function.
In Bi$_2$Se$_3$ and Bi$_2$Te$_3$ there is a single Weyl-like fermion species \cite{Zhang09,Xia09,Y_Chen09}) 
which can be described by an effective 2D Hamiltonian \cite{Fu09,Liu10}:
\begin{equation}
H = {\cal A} (\sigma_x k_y - \sigma_y k_x) + {\cal W}_{\bf k} \sigma_z +  {\cal D}{\bf k}^2 \sigma_0, 
\label{H_BiSe_0}
\end{equation} 
\begin{equation}
{\cal W}_{\bf k }=\frac{W}{2}( k^3_+ + k^3_-), 
\qquad 
k_\pm = k_x \pm  ik_y, 
\label{W}
\end{equation}
where ${\bf k} = (k_x, k_y)$ is the wave vector on the surface and 
${\cal A}$, ${\cal D}$ and ${\cal W}$ are band structure parameters. 
Upon unitary transformation $H \to U H U^\dagger$ with
$
U = \bigl(
  \begin{smallmatrix}
    i & 0 \\
    0 & 1 \\
  \end{smallmatrix}
\bigr),
$
Hamiltonian (\ref{H_BiSe_0}) takes the form:
\begin{equation}
H = \mbox{\boldmath$\sigma$} \cdot ( {\cal A} {\bf k} + {\cal W}_{\bf k} {\bf z} ) +  
{\cal D}{\bf k}^2 \sigma_0, 
\label{H_BiSe}
\end{equation} 
which looks similar to the HgTe quantum well Hamiltonian (\ref{H_HgTe_D}). 
There are, however, two important distinctions between these Hamiltonians. 
First, here the basis functions correspond to $\frac{1}{2}$ and $-\frac{1}{2}$ electron spin projections, 
i.e. Pauli ($\sigma_{x,y,z}$) and unit ($\sigma_0$) matrices act on real spin indices. 
Second, the ${\cal W}_{\bf\hat k}$-term in Eq.~(\ref{H_BiSe}) is cubic (odd) in momentum 
${\bf\hat k}$, causing no gap at ${\bf k}=0$. This term does not break 
${\bf k},\mbox{\boldmath$\sigma$} \to -{\bf k},-\mbox{\boldmath$\sigma$}$ invariance,
which is the real time-reversal symmetry in this case. 
Instead, it causes hexagonal warping \cite{Alpichshev10,Kuroda10b} of the surface spectrum 
(see, also Figs.~\ref{E_3D}): 
\begin{eqnarray}
E(k,\phi_{\bf n})=
{\cal D}k^2 \pm \sqrt{  {\cal A}^2k^2 + \frac{W^2k^6}{2}( 1 + \cos 6\phi_{\bf n})  },
 \label{Spectrum_BiSe}
\end{eqnarray}
where the angle $\phi_{\bf n}$ indicates the momentum direction.

\subsection{Persistence of surface spin-momentum locking in the presence of disorder.}
\label{Persistance}

In Eqs. (\ref{H_BiSe_0}) and (\ref{H_BiSe}) the first linear term is analogous to the Bychkov-Rashba 
spin-orbit interaction that occurs in a conventional 2D electronic system confined in an asymmetric quantum well \cite{Bychkov84}. 
The key difference here is the strength of the coupling constant ${\cal A}$. For the topological surface states, the experimentally determined 
coupling constants typically are ${\cal A} = 355$ meV$\cdot$nm for Bi$_2$Se$_3$ \cite{Kuroda10b} 
and ${\cal A} = 255$ meV$\cdot$nm for Bi$_2$Te$_3$ \cite{Y_Chen09,Alpichshev10}. 
These values are really giant compared to the Bychkov-Rashba coupling 
$\alpha_{BR} \sim 1 \div 10$ meV$\cdot$nm in conventional GaAs-based 2D electronic systems (see e.g. Ref. \cite{Zutic04}). 
This large quantitative difference has two important physical consequences. 
The first, as mentioned above, is the formation of the conical band dispersion [see also Fig. \ref{E_3D}]. 
The second is the robustness of the surface spin-momentum locking against potential impurity scattering (and other spin-independent scattering) 
as long as the energy separation between the valence and conduction bands [see Eq. (\ref{Spectrum_BiSe})] is larger 
than the spectrum broadening $\hbar/\tau_0$:
\begin{equation}
2\sqrt{ {\cal A}^2k^2_F + {\cal W}^2_{ {\bf k}_F } } = 
2(E_F - {\cal D}k^2_F) \gg \hbar/\tau_0,
 \label{weak}
\end{equation}
where $\tau_0$ is the elastic life-time, $k_F$ and $E_F$ is the Fermi momentum and energy 
(the Fermi level is, for concreteness, located in the conduction band). 
Due to the large constant ${\cal A}$ the condition (\ref{weak}) 
can be met already for the Fermi energy values, $E_F$, of several tens of meV. 
Moreover, the condition (\ref{weak}) simultaneously implies a diffusive metallic transport regime in which   
\begin{equation}
2E_F \, \tau_0/\hbar \approx k_F v_F \tau = k_F \ell \gg 1, 
\label{diff}
\end{equation}
because for a conical dispersion the transport momentum relaxation time 
$\tau \approx 2\tau_0$ \cite{Ando05} ($v_F$ is the Fermi velocity and 
$\ell=v_F \tau$ is the transport mean-free path). 
Thus, the large values of the coupling constant ${\cal A}$ [see Eq. (\ref{weak})] 
enable the diffusive metallic regime (\ref{diff}) without significant spin relaxation, 
i.e. the spin-helicity of the carriers is preserved during their diffusion on a disordered TI surface. 
For Hamiltonian (\ref{H_BiSe}) the conservation of the spin-helicity can be expressed as 
\begin{equation}
\mbox{\boldmath$\sigma$}\cdot\frac{ {\cal A} {\bf k} + {\cal W}_{\bf k} {\bf z} }
{|{\cal A} {\bf k} + {\cal W}_{\bf k} {\bf z}|} = h = {\rm const},
\label{helicity}
\end{equation}
where the constant $h$ takes the values $\pm 1$ for the conduction and valence bands, respectively.
The absence of the spin relaxation under condition (\ref{helicity}) defines a distinct transport regime - the 2D helical metal -,
opening hitherto unexplored routes for theoretical investigations of disordered systems 
(e.g. Refs. \cite{Ostrovsky10,Culcer10,Culcer11a,Culcer11b,Culcer12}). 

\begin{figure}[t]
\begin{center}
\includegraphics[width=85mm]{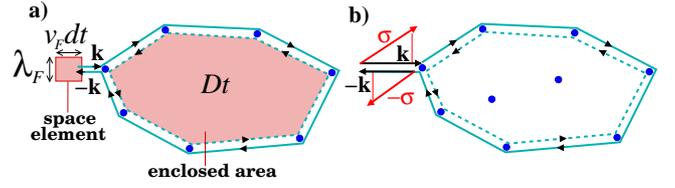}
\end{center}
\caption{
(a) Schematic of closed electron trajectories giving rise to weak localization [see also Eq. (\ref{WL_est}) and text]. 
Shaded areas indicate a space element of the trajectory, $\lambda_F v_F dt$, and the area enclosed by the trajectory, $Dt$,
which we use to estimate the return probabilty $dP_{ret}(t)$ in Eq. (\ref{WL_est}). 
(b) Same trajectories on the surface of a 3D TI involve opposite momenta, ${\bf k}$ and ${\bf -k}$, and 
opposite spins, $\mbox{\boldmath$\sigma$}$ and $-\mbox{\boldmath$\sigma$}$, 
as a result of the conservation of the spin-helicity [see also Eq. (\ref{h_cons}) and text]. 
This leads to the weak antilocalization conductivity (\ref{WAL_res}). 
}
\label{WL}
\end{figure}

\subsection{ Surface weak antilocalization in 3D TIs.}
\label{Surface}

One important example of observable transport phenomena in disordered systems is the weak localization (WL) effect 
\cite{Altshuler80,Bergmann84,Akk_Mont07}. 
It is associated with closed classical electron trajectories that self intersect with opposite electron momenta, 
${\bf k}$ and ${\bf -k}$ (see Fig. \ref{WL}a). On such trajectories, two states which are related to each other by time reversal 
interfere constructively, yielding, in the absence of the spin-orbit coupling, a negative WL correction $\delta\sigma$ 
to the classical Drude conductivity $\sigma_D$. The relative change $\delta\sigma/\sigma_D$ is the measure of the classical return probability 
integrated from the shortest diffusion time-scale $\tau$ to the largest one, given by the dephasing time $\tau_\varphi$: 
\begin{equation}
\frac{ \delta\sigma }{ \sigma_D } \sim - \int_\tau^{\tau_\varphi} dP_{ret}(t), \,\, 
dP_{ret}(t) \sim \frac{\lambda_F v_F dt}{ (\sqrt{Dt})^2}.
\label{WL_est}
\end{equation}
The negative sign here is the consequence of the fact that the particle returns to its original location with the opposite momentum ${\bf -k}$ 
[i.e. the Kubo formula for the conductivity correction contains the product of the two opposite-sign velocities]. 
The return probability $dP_{ret}(t)$ is proportional to a space element of the trajectory, $\lambda_F v_F dt$, 
divided by the typical area $(\sqrt{Dt})^2$ enclosed by the particle trajectory at time $t$ 
($\lambda_F$ is the Fermi wavelength and $D$ is the diffusion constant). 
The conductivity correction (\ref{WL_est}) is logarithmically divergent and has a universal prefactor 
\cite{Altshuler80,Bergmann84,Akk_Mont07}: 
\begin{equation}
\delta\sigma = - \frac{2e^2}{2\pi h}\ln\frac{\tau_\varphi}{\tau},
\label{WL_res}
\end{equation}
where the factor of 2 in the numerator is due to the spin degeneracy.

In contrast to Eq. (\ref{WL_res}), according to experiments 
\cite{Checkelsky09,J_Chen10,Checkelsky11,He11,Bouvier11,Steinberg11,Kim11,J_Chen11,Cha12,Takagaki12,Takagaki12_b} 
and theoretical calculations \cite{GT11_WAL,Nestoklon11,Lu11a,Lu11b,Adroguer12,Garate12}, 
the topological surface states in 3D TIs exhibit the weak antilocalization (WAL) effect characterized 
by a positive conductivity correction $\delta\sigma > 0$. 
The change of the sign can be easily explained by the fact that the classical motion along the loop 
is now subject to the conservation of the spin-helicity [see Eq. (\ref{helicity})]. Indeed, upon returning to its original location 
the particle with momentum ${\bf -k}$ must have the same helicity as it had initially with momentum ${\bf k}$. 
This requires the change of the direction of the electron spin from $\mbox{\boldmath$\sigma$}$ to $-\mbox{\boldmath$\sigma$}$ (see also Fig. \ref{WL}b)
so that we have the identity
\begin{equation}
\mbox{\boldmath$\sigma$}\cdot\frac{ {\cal A} {\bf k} + {\cal W}_{\bf k} {\bf z} }
{|{\cal A} {\bf k} + {\cal W}_{\bf k} {\bf z}|} 
=
-\mbox{\boldmath$\sigma$}\cdot\frac{ -{\cal A} {\bf k} + {\cal W}_{-\bf k} {\bf z} }
{|{\cal A} {\bf k} + {\cal W}_{\bf k} {\bf z}|}. 
\label{h_cons}
\end{equation} 
The spin rotation $\mbox{\boldmath$\sigma$} \to -\mbox{\boldmath$\sigma$}$ yields another "minus" sign in Eq. (\ref{WL_est}):
\begin{equation}
\delta\sigma = + \frac{e^2}{2\pi h}\ln\frac{\tau_\varphi}{\tau}.
\label{WAL_res}
\end{equation}
This can also be viewed as the result of the $\pi$ Berry phase accumulated along the trajectory loop \cite{Ando05,Suzuura02}. 

The explicit calculation of $\delta\sigma$ with the account of the warping ${\cal W}_{\bf k}$ under condition (\ref{weak}) 
was carried out in Refs. \cite{GT11_WAL,Adroguer12}. We note that the conductivity correction $\delta\sigma$ (\ref{WAL_res}) has the same form 
as for a conventional 2D electron system with spin-orbit impurity scattering \cite{Hikami80} or with Bychkov-Rashba and Dresselhaus 
spin-orbit interactions (see e.g. Refs. \cite{Iordanskii94,Knap96}). The reason is that the surface states of 3DTIs with hexagonal warping
and conventional 2D electron systems with spin-orbit impurity scattering, Bychkov-Rashba or Dresselhaus 
spin-orbit interactions belong to the same - symplectic - universality class of disordered systems. 

\begin{figure}[t]
\begin{center}
\includegraphics[width=70mm]{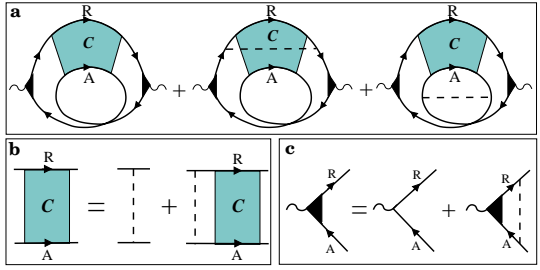}
\end{center}
\caption{
Diagrammatic representations for
(a) bare and dressed Hikami boxes for the correction to Drude conductivity, 
(b) Bethe-Salpeter equation for the Cooperon, and 
(c) equation for the renormalized current vertex in the ladder approximation. 
Thick lines denote disorder-averaged Green's functions in self-consistent Born approximation, dashed lines - the potential (spin-independent) 
disorder correlation functions.
}
\label{Diagrams}
\end{figure}

In contradiction with the aforementioned symmetry argument, Ref. \cite{Wang11} has found a nonuniversal prefactor in Eq. (\ref{WAL_res}) 
which depends on the hexagonal warping strength. It is therefore worthwhile to briefly review the calculation idea and show   
that different warping terms cancel each other, yielding a universal prefactor $e^2/2\pi h$ in Eq. (\ref{WAL_res}).   
We take the standard formulas for the quantum-interference conductivity correction, which are expressed diagrammatically 
in Fig. \ref{Diagrams} (see e.g. Refs. \cite{Akk_Mont07}, \cite{McCann06}), 
treating the warping (\ref{W}) as weak perturbation onto the isotropic spectrum under condition 
\begin{equation}
W^2k_F^4 / 2 {\cal A}^2 \ll 1.
\label{w}
\end{equation}
The first diagram in Fig. \ref{Diagrams}a - the standard bare Hikami box - gives the following result:
\begin{equation}
\delta\sigma^{bare}=
\frac{ \tau }{ \tau_0 } \times \frac{e^2}{2\pi h}\ln\frac{\tau_\varphi}{\tau}.
\label{dS_bare}
\end{equation}
The nonuniversal prefactor $\tau /\tau_0 $ reflects
the difference between the elastic life-time $\tau_0$ and transport relaxation time $\tau$  
for helical carriers \cite{GT11_WAL}:
\begin{equation}
\frac{ \tau }{\tau_0}= \frac{2}{ 1 + \overline{ W^2_{ {\bf k}_F } }/{\cal A}^2k^2_F } = 
\frac{2}{ 1 + W^2k_F^4/2{\cal A}^2 }, 
\label{tau_tr}
\end{equation}
formally described by the vertex renormalization in Fig. \ref{Diagrams}c. 
In Eq. (\ref{tau_tr}) the bar denotes averaging over the directions of the momentum, ${\bf k}_F$, on the Fermi surface. 
Since $\tau /\tau_0 \not = 1$, two other diagrams in Fig. \ref{Diagrams}a - dressed Hikami boxes -  
also need to be taken into account. Each of them gives the following correction:
\begin{equation}
\delta\sigma^{dres} =
-\frac{1 - \tau_0 /\tau}{2} \delta\sigma^{bare}.
\label{dS_dressed}
\end{equation}
For the unwarped Dirac cone ($\tau = 2\tau_0$), $\delta\sigma^{dres} = (-1/4) \times \delta\sigma^{bare}$, 
in agreement with the calculations of the WL in graphene \cite{McCann06}. 
Thus, the net result is 
\begin{eqnarray}
\delta\sigma = \delta\sigma^{bare} + 2\delta\sigma^{dres} 
= \frac{ \tau_0 }{ \tau } \,\, \delta\sigma^{bare} = \frac{e^2}{2\pi h}\ln\frac{\tau_\varphi}{\tau}.
\nonumber
\end{eqnarray}

\subsection{ Detecting surface states by WAL magnetotransport. }
\label{Extracting}

So far we have treated the surface states in 3D TIs as purely two-dimensional. 
There is, however, a finite length $\lambda$ of order of a few nm over which they penetrate into the bulk of the material. 
Since $\lambda$ is quite small, the surface WAL conductivity is sensitive to the orientation of an external magnetic field 
with respect to the surface of the material, which could be used in practice to detect the surface states 
(see e.g. recent experiment \cite{He11}). Defining the magnetoconductivity as $\Delta\sigma(B)=\delta\sigma(B)-\delta\sigma(0)$ , 
one can obtain the following $B$-field dependences for perpendicular ($\perp$) and parallel ($\|$) field orientations:
\begin{equation}
\Delta\sigma_\perp (B)=\frac{e^2}{2\pi h}
\Biggl[
\ln\frac{ B_\perp }{B}
-\psi\Biggl(
\frac{1}{2} + \frac{ B_\perp }{B}
\Biggr)
\Biggr],
\,
B_\perp=\frac{ \hbar }{ 4|e| \, \ell^2_\varphi },
\label{Out}
\end{equation}
\begin{equation}
\Delta\sigma_\| (B) = -\frac{e^2}{2\pi h}\ln\Biggl( 1 + \frac{B^2}{ B^2_{_\|} } \Biggr), \quad
B_{_\|} = \frac{\hbar}{\sqrt{2} |e| \lambda \ell_\varphi }.
\label{In}
\end{equation}
These equations contain the same phase-coherence length $\ell_\varphi = \sqrt{ D\tau_\varphi }$. 
However, the magnetic-field scales, $B_\perp$ and $B_{_\|}$, on which the magnetoconductivity decreases, are distinctly different. 
The field $B_\perp$ corresponds to the Aharonov-Bohm magnetic flux of order of the quantum $h/e$  
through a typical area $\sim \ell^2_\varphi$ enclosed by the interfering trajectories \cite{Altshuler80}, 
while $B_\|$ corresponds to the same flux $h/e$, but through a significantly smaller area 
$\sim \lambda \ell_\varphi$ which is proportional to the surface-state penetration length in the bulk \cite{B_par}.   

In order to identify the surface state one should extract its penetration length $\lambda$ from Eqs. (\ref{Out}) and (\ref{In}).
Excluding $\ell_\varphi$ from Eqs. (\ref{Out}) and (\ref{In}) we can express $\lambda$ in terms of two parameters, 
$B_\perp$ and $B_{_\|}$, which can be obtained independently from fitting the corresponding experimental data \cite{GT11_WAL}:
\begin{equation}
\lambda = \sqrt{ \frac{2\hbar}{|e|} \frac{ B_\perp }{ B^2_{_\|} } }.
 \label{lambda}
\end{equation}
The knowledge of this penetration length also helps to estimate the critical thickness of the sample 
at which the two surface states start to overlap and the TI state disappears.  

To conclude this section, we should mention that the WAL in 3DTIs is suppressed and may even turn into the WL if an energy gap opens in the surface spectrum,
as a result of time-reversal symmetry breaking \cite{Lu11a}. This should apply to 3DTIs with magnetically doped surfaces (see e.g. Ref. \cite{He11}). 
The WL behavior has also been found in ultrathin TI films in which the lowest bulk-state subbands are described by a massive 2D Dirac model \cite{Lu11b}. 
Besides, quantum transport in electrically gated TI films may depend sensitively on the coupling between surface and bulk states \cite{Garate12}. 
Such a coupling introduces new time scales on which the crossover from WAL to WL may occur \cite{Garate12}.

\section{Quantum Hall effect on topological surfaces.}
\label{3DTIs_QHE}
Another manifestation of the surface spin-helicity in 3DTIs is the unconventional 
half-integer quantum Hall effect (QHE)  (see e.g. Refs. \cite{Fu07b,Qi08,Tse10a,Tse10b,Tse11}). 
Such unconventional QHE was first observed in single atomic layers of carbon - graphene - \cite{Neto09}, and has been regarded as a purely 2D phenomenon. 
The search for the half-integer Hall quantization in 3DTI systems is therefore a new challenging task 
(see e.g. Refs. \cite{Hasan10,Tse10a,Tse10b,Tse11,Bruene11}). 
Our discussion of the QHE in 3DTIs follows closely Refs. \cite{Tse10b,Tse11}. 

\begin{figure}[t]%
\begin{center}
\includegraphics[width=40mm]{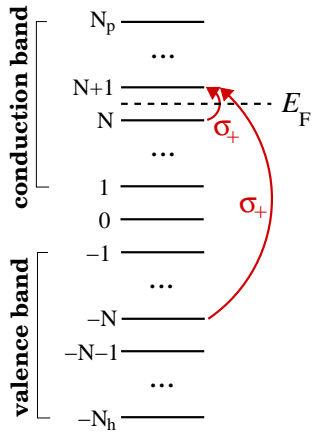}
\end{center}
\caption{%
Schematic of surface Landau level spectrum (\ref{LL_n}). 
For particle-hole symmetric case, only two dipole transitions $N\to N+1$ (intraband) and $-N\to N+1$ (interband) contribute to Hall conductivity, 
yielding the half-integer quantized $\sigma_{xy} = (e^2/h)(N + 1/2)$ (\ref{s_xy_dc}); 
$N$ is the index of the highest occupied Landau level (\ref{N}), $E_F$ is the Fermi level.
}
\label{LLs_sym}
\end{figure}

\subsection{Landau quantization of surface states and conductivity tensor.}
Using a linear Hamiltonian 
$H =  v \mbox{\boldmath$\sigma$}\cdot ( -i\hbar \nabla - e{\bf A} ) $ 
for a single topological surface state in a perpendicular magnetic field (${\bf A}$ is the vector potential), we arrive at the following eigenvalue problem:
\begin{eqnarray}
H \overline{ | n }\rangle  = \epsilon \overline{ | n} \rangle ,\,\, 
H= 
\left[
\begin{array}{cc}
0   &  -i\sqrt{2}\hbar\Omega_B\, a    \\
i\sqrt{2}\hbar\Omega_B\, a^\dagger  &  0 
\end{array}
\right],
 \label{Eigen}
\end{eqnarray}
where $\overline{ | ... }\rangle$ denotes an eigenspinor, 
$a^\dagger$ and $a$ are the raising and lowering operators of a harmonic oscillator, respectively, 
and $\hbar \Omega_B = \hbar v/\ell_B$ is the characteristic Landau level (LL) spacing depending on the magnetic length $\ell_B = \sqrt{ \hbar/|eB| }$. 
The solutions for the LLs and eigenspinors are given by 
\begin{eqnarray}
&&
\epsilon_n = {\rm sgn}(n) \hbar \Omega_B \sqrt{2 |n| }, \,
\overline{ | n }\rangle = 
\left[
\begin{array}{c}
-iC_{\uparrow n}| |n|-1 \rangle   \\
C_{\downarrow n}| |n|   \rangle 
\end{array}
\right],
 \label{LL_n}\\
%
&&
C_{\uparrow n} = {\rm sgn}(n) /\sqrt{ 2 } , \quad 
C_{\downarrow n} = 1 / \sqrt{ 2 } , \quad n\not =0,
\label{C}\\
%
&&
\epsilon_0 = 0, \quad 
\overline{ | 0 }\rangle = 
\left[
\begin{array}{c}
0   \\
| 0   \rangle 
\end{array}
\right],\quad 
C_{\uparrow 0} =0, \, C_{\downarrow 0} =1, \quad n=0. 
\nonumber
\end{eqnarray}
In the LL basis the Kubo conductivity tensor is
\begin{equation}
\sigma_{\alpha\beta}(\omega) = \frac{ i\hbar }{ 2\pi\ell^2_B }\sum_{n,n^\prime}
\frac{ f_n - f_{n^\prime} }{ \epsilon_n - \epsilon_{n^\prime} }
\frac{
\langle \overline{n|}j_\alpha\overline{|n^\prime}\rangle \langle \overline{ n^\prime|}j_\beta\overline{|n }\rangle
}
{
\hbar\omega + \epsilon_n - \epsilon_{n^\prime} + i\hbar/\tau
},
\label{Kubo}
\end{equation}
where $j_{\alpha (\beta)} = ev \sigma_{\alpha (\beta) }$ is the $\alpha (\beta)$-component of the surface current operator 
($\alpha,\beta=x,y$), and $f_n$ is the Fermi occupation number of the nth LL. 
Introducing operators $\sigma_\pm = (\sigma_x \pm i\sigma_y)/2$ we find that the current matrix elements  
$
\langle \overline{n^\prime|}j_x\overline{|n}\rangle = 
ev [ \langle \overline{n^\prime|} \sigma_- \overline{|n}\rangle + \langle \overline{n^\prime|} \sigma_+ \overline{|n}\rangle ] 
$
and 
$
\langle \overline{n^\prime|}j_y\overline{|n }\rangle = 
iev [ \langle \overline{n^\prime|} \sigma_- \overline{|n}\rangle - \langle \overline{n^\prime|} \sigma_+ \overline{|n}\rangle ],
$
obey the dipole selection rules 
\begin{equation}
\langle \overline{n^\prime|} \sigma_\pm \overline{|n}\rangle = \pm i C_{\downarrow (\uparrow) n}C_{\uparrow (\downarrow) n^\prime} \delta_{|n^\prime|,|n| \pm 1},
\label{s_pm}
\end{equation}
allowing only transitions between LLs with $|n^\prime|=|n| \pm 1$. Below we illustrate a link between selection rules (\ref{s_pm}) 
and the half-integer quantization of the Hall conductivity.

\subsection{Half-integer-quantized Hall conductivity $\sigma_{xy}$. Role of particle-hole symmetry.}

The unconventional half-integer-quantized Hall conductivity of the helical carriers 
is linked to their Berry phases. One way to see this is to calculate the topological Chern number associated with the Berry flux 
in the Brillouin zone, as discussed in detail, e.g., in Refs. \cite{Hasan10,Qi11,Qi08}. 
Here, we intend to obtain this result directly from Kubo formula (\ref{Kubo}), 
as this allows us to address a more general situation in which particle-hole asymmetry and finite-frequency effects also play a role.  

We are interested in the QHE regime that requires strong magnetic fields such that 
\begin{equation}  
\sqrt{2}\Omega_B\tau =4R_0 \sigma_s \sqrt{ \frac{|B|}{n_s \Phi_0} } \gg 1.
\label{QHE_regime}
\end{equation}
The realization of this regime depends also on the classical surface conductivity 
$\sigma_s = (e^2/h) (k_F\ell/2)$ and surface carrier density $n_s$ 
($R_0 = h/2e^2$ and $\Phi_0 = h/|e|$ are the resistance and magnetic flux quanta).
Additionally, we consider the zero temperature limit of conductivity (\ref{Kubo}) 
in which all LLs below the Fermi level $E_F$ are occupied,  
\begin{equation}
-N_h \leq n \leq N, \qquad f_n =1,
\label{Occ}
\end{equation}
while those above $E_F$ are empty (see also Fig. \ref{LLs_sym}):
\begin{equation}
N + 1 \leq n \leq N_p, \qquad f_n =0, 
\label{Emp}
\end{equation}
where $N$ is the index of the highest occupied LL given by
\begin{equation}
N = {\rm Int}\biggl(\frac{E_F^2}{2\hbar^2\Omega^2_B} \biggr) = {\rm Int} \biggl( 2\pi n_s \ell^2_B \biggr) 
= {\rm Int}\biggl( \frac{n_s \Phi_0}{|B|} \biggr),
\label{N}
\end{equation}
where ${\rm Int}(...)$ denotes the integer part. In Eqs. (\ref{Occ}) and (\ref{Emp}) 
we introduce the number of the hole LLs in the valence band, $N_h$, 
and the number of the particle LLs in the valence band, $N_p$, both counted from the charge neutrality point. 
Using Eqs. (\ref{Occ}), (\ref{Emp}) and selection rules (\ref{s_pm}) we can write the Hall conductivity (\ref{Kubo}) as  
\begin{eqnarray}
&&
\sigma_{xy}(\omega) = \frac{ e^2  }{ h } 2\Omega^2_B 
\left[
\frac{
1/4
}
{
( \epsilon_{N+1} - \epsilon_{N} )^2/\hbar^2 + (\tau^{-1} - i\omega)^2 
}
\right.
\nonumber\\
&&
+
\frac{
1/4
}
{
( \epsilon_{N+1} - \epsilon_{-N} )^2/\hbar^2 + (\tau^{-1} - i\omega)^2 
}
\label{s1_xy}\\
&&
\left.
+
\sum\limits_{n^\prime=N+1}^{N_p}\sum\limits_{n=N+1}^{N_h}
\frac{
| \langle \overline{n^\prime|}\sigma_+\overline{|n }\rangle|^2 - |\langle \overline{ n|}\sigma_+\overline{|n^\prime }\rangle|^2
}
{
( \epsilon_{n^\prime} + \epsilon_n )^2/\hbar^2 + (\tau^{-1} - i\omega)^2 
}
\right].
\label{s2_xy}
\end{eqnarray}
\begin{figure}[t]%
\begin{center}  
\includegraphics[width=0.6\linewidth]{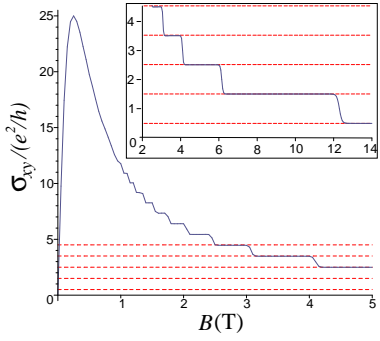}%
\end{center}
  \caption{%
DC Hall conductivity $\sigma_{xy}$ [given by Eq. (\ref{Kubo}) for $\omega=0$] in units of $e^2/h$ versus magnetic field $B$ in Tesla. 
Inset: same dependence in high-field region; surface state parameters are $R_0\sigma_s = 25$, $n_s=3\cdot 10^{11}$cm$^{-2}$, and
$\hbar v = 330$ meV$\cdot$nm, and temperature $T=2$K.    
}
\label{s_xy_B}
\end{figure}
In this equation the first term corresponds to an intraband transition between the LLs $N$ and $N +1$, both in the conduction band,
whereas the second term (\ref{s1_xy}) corresponds to an interband transition from the occupied valence-band LL $-N$ to the empty LL $N+1$ 
in the conduction band (see also Fig. \ref{LLs_sym}). The third term (\ref{s2_xy}) involves interband transitions from the rest of the occupied (hole) 
LLs $-N_h \leq n \leq -N-1$ to empty (particle) LLs $N+1\leq n^\prime \leq N_p$. 
Since the summand in Eq. (\ref{s2_xy}) is odd under exchange $n \leftrightarrow n^\prime$, this term vanishes for the particle-hole symmetric spectrum 
with $N_p=N_h$. This means that the interband transitions induced by the spin-raising ($\sigma_+$) and -lowering  ($\sigma_-$) operators between 
the particle-hole symmetric LLs $n \leq -N-1$ and $n^\prime \geq N+1$  exactly compensate each other. 
Therefore, in the QHE regime (\ref{QHE_regime}) the dc conductivity $\sigma_{xy} \equiv \sigma_{xy}(\omega=0)$ is 
\begin{eqnarray}
\sigma_{xy} &=& \frac{ e^2  }{ h }
\left[
\frac{1/4}
{
( \sqrt{N+1} - \sqrt{N} )^2  
}
+
\frac{1/4}
{
( \sqrt{N+1} + \sqrt{N} )^2  
}
\right]=
\nonumber\\
&=& \frac{ e^2  }{ h } \left[ N + \frac{1}{2} \right], \qquad N_p = N_h.
\label{s_xy_dc}
\end{eqnarray}
Thus, the half-integer Hall conductivity (\ref{s_xy_dc}) reflects the particle-hole symmetry in two ways. 
First, it involves the transitions from the symmetric particle and hole LLs, $N$ and $-N$, to the lowest empty LL $N+1$, and, second, 
interband transitions from deep LLs $-N_h,...,-N-1$ to higher empty LLs $N+1,...,N_p$ do not affect $\sigma_{xy}$ 
provided that the conduction and valence bands contain equal LL numbers, $N_p=N_h$. 
In Fig. \ref{s_xy_B} we plot the dc Hall conductivity $\sigma_{xy}$ in the whole magnetic field region and at finite temperature $T$, 
using Eq. (\ref{Kubo}) with $\omega=0$.

\subsection{Particle-hole asymmetry $N_p \not = N_h$. Accuracy of the half-integer quantization.}

\begin{figure}[t]%
\begin{center}
\includegraphics*[width=60mm]{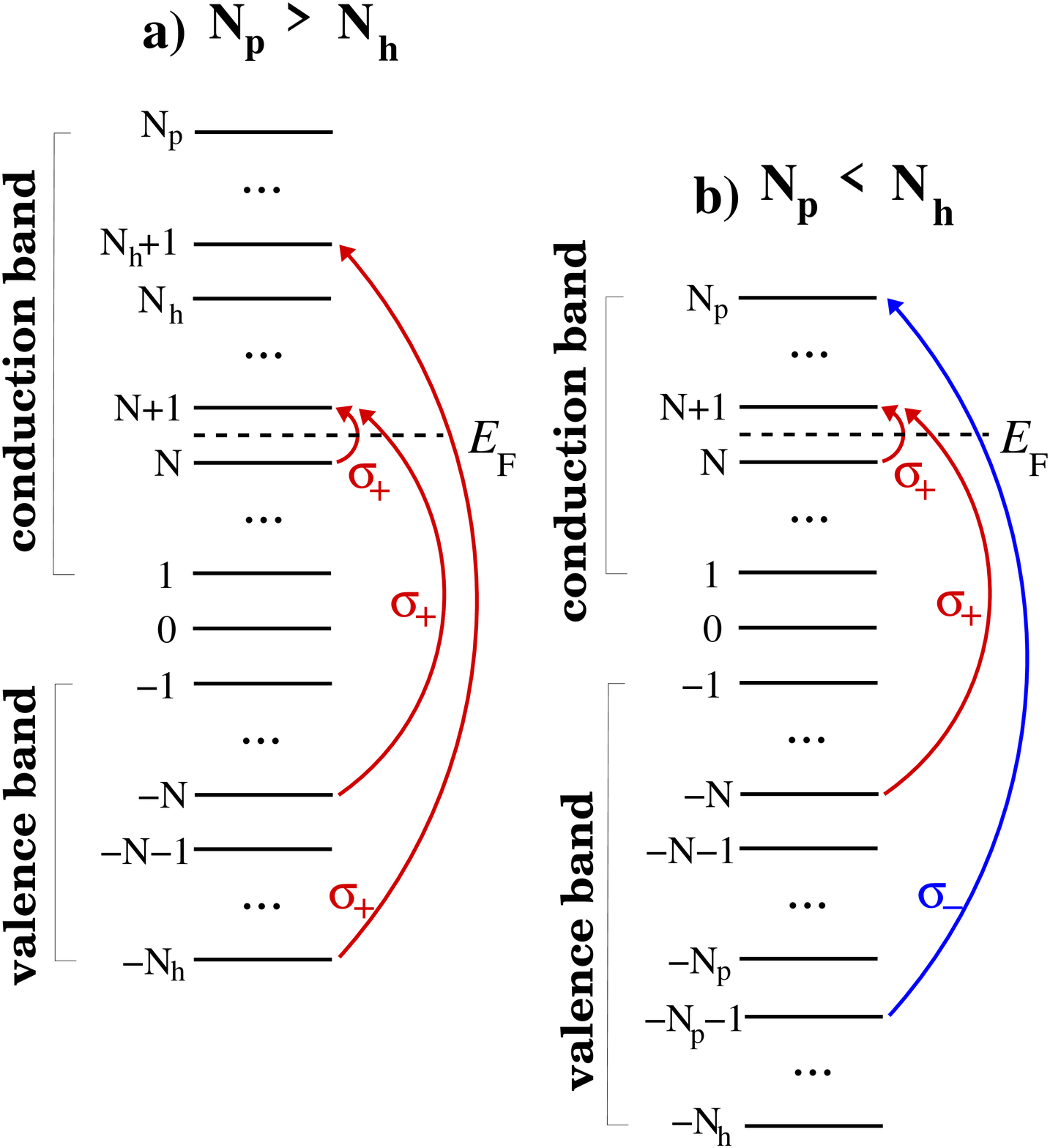}
\end{center}
\caption{%
Particle-hole asymmetric Landau spectrum with (a) $N_p > N_h$ and (b) $N_p < N_h$. 
Additional allowed interband transitions $-N_h\to N_h+1$ (a) and $-N_p-1\to N_p$ (b) give rise to non-universal corrections 
to the half-integer quantized Hall conductivity [see also Eqs. (\ref{s_xy_p}), (\ref{s_xy_h}), and (\ref{s_xy_Corr})].
}
\label{LLs_asym}
\end{figure}

We now proceed by discussing the Hall conductivity for the particle-hole asymmetric LL spectrum with $N_p > N_h$. 
In this case $\sigma_{xy}$ acquires an additional contibution due to the interband transitions in Eq. (\ref{s2_xy}),  
\begin{eqnarray}
&&
\sigma_{xy} = \frac{ e^2  }{ h } 
\left[
 N + \frac{1}{2} + 2(\hbar\Omega_B)^2 \times
\right.
\label{s5_xy}\\
&&
\left.
\times \sum\limits_{n^\prime=N_h +1 }^{N_p}\sum\limits_{n = -N_h}^{ -N - 1 }
\frac{
| \langle \overline{n^\prime|}\sigma_+\overline{|n }\rangle|^2 - |\langle \overline{ n^\prime|}\sigma_-\overline{|n }\rangle|^2
}
{
( \epsilon_{n^\prime} - \epsilon_n )^2  
}
\right].
\label{s6_xy}
\end{eqnarray}
The last term contains an additional allowed transition from the deepest hole LL $-N_h$ to the empty particle state $N_h+1$ 
(see also Fig. \ref{LLs_asym}a) with matrix element $\langle \overline{N_h + 1|}\sigma_+\overline{| -N_h }\rangle$. 
Therefore, for $N_p > N_h$ we have
\begin{eqnarray}
&&
\sigma_{xy} = \frac{ e^2  }{ h } 
\left[
 N + \frac{1}{2} + \frac{1/4}{( \sqrt{N_h + 1} + \sqrt{N_h} )^2}
\right].
\label{s_xy_p}
\end{eqnarray}
In the similar way we obtain $\sigma_{xy}$ for the particle-hole asymmetric LL spectrum with $N_p < N_h$: 
\begin{eqnarray}
&&
\sigma_{xy} = \frac{ e^2  }{ h } 
\left[
 N + \frac{1}{2} - \frac{1/4}{( \sqrt{N_p} + \sqrt{N_p + 1} )^2}
\right].
\label{s_xy_h}
\end{eqnarray}
In this case there is an additional allowed transition from the hole LL $-N_p-1$ to the highest empty particle state $N_p$ induced by 
the spin-lowering operator $\sigma_-$ (see also Fig. \ref{LLs_asym}b), due to which the correction to the half-integer $N+1/2$ is negative.  
To summarize, for large $N_p, N_h \gg 1$ the Hall conductivity is given by  
\begin{eqnarray}
\sigma_{xy} = \frac{ e^2  }{ h } 
\left[
 N + \frac{1}{2} + \delta
\right], 
\quad 
\delta \approx \frac{ {\rm sgn}(N_p - N_h) }{ 16 \cdot {\rm min}(N_p,N_h) },  
\label{s_xy_Corr}
\end{eqnarray}
where correction $\delta$ determines the accuracy of the half-integer quantization.
\begin{figure}[t]%
\begin{center}  
\includegraphics[width=0.6\linewidth]{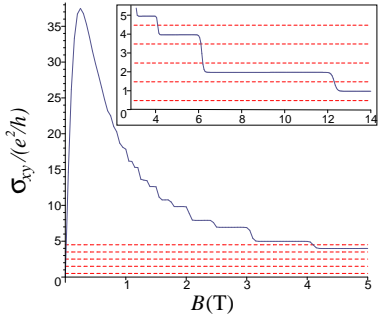}%
\end{center}
  \caption{%
Two-surface DC Hall conductivity $\sigma_{xy}$ [see Eqs. (\ref{Kubo}) and (\ref{2s_xy}) for $\omega=0$] 
in units of $e^2/h$ versus magnetic field $B$ in Tesla. Inset: same dependence in high-field region; 
top (t) and bottom (b) surface state parameters are $R_0\sigma^t_s = 25$, $n^t_s=3\cdot 10^{11}$cm$^{-2}$ and 
$R_0\sigma^b_s = 12.5$, $n^b_s=1.5\cdot 10^{11}$cm$^{-2}$. For both surfaces 
$\hbar v = 330$ meV$\cdot$nm and $T=2$K.    
}
\label{2s_xy_B}
\end{figure}

\subsection{Two-surface model.}
\label{2surface}

Since in a 3DTI two surface states, one at the top and one at the bottom of the sample, 
can contribute to transport \cite{Bruene11}, it is worthwhile to briefly discuss the behavior of the net Hall conductivity:
\begin{equation}
 \sigma_{xy} = \sigma^{t}_{xy} + \sigma^{b}_{xy},
\label{2s_xy}
\end{equation}
where the conductivities of the top and bottom surfaces, $\sigma^{t,b}_{xy}$,
are both given by Eq. (\ref{Kubo}) with, generally, different classical conductivities $\sigma^{t,b}_s$ and carrier densities $n^{t,b}_s$.
Such difference may arise when one surface faces a substrate, while the other - vaccuum \cite{Bruene11}.
Although, separately, each surface exhibits the half-integer Hall quantization, the plateaus in the net conductivity 
deviate from $(e^2/h)(N+1/2)$ and are irregular, e.g. Fig. \ref{2s_xy_B} shows an unusual plateau sequence $e^2/h, 2e^2/h, 4e^2/h, 5e^2/h$...  
The QHE with unusual odd and even plateaus has been observed in the 3DTI HgTe where a bulk energy gap is induced by epitaxial strain \cite{Bruene11}.   
The bulk energy gap is a necessary prerequisite for the observation of the surface contribution in transport which is, otherwise, dominated by bulk carriers 
\cite{Taskin10,Butch10,Eto10,Analytis10}.

\subsection{AC conductivities. Classical cyclotron resonance.}
\label{AC}

Here we briefly discuss the AC Hall $\sigma_{xy}(\omega)$ and diagonal $\sigma_{xx}(\omega)$ conductivities. 
At zero temperature $T=0$ for the particle-hole symmetric LL spectrum, they can be obtained from the Kubo formula (\ref{Kubo}) as
\begin{eqnarray}
&&
\sigma_{xy}(\omega) = \frac{ e^2  }{ h } 
\Biggl[
\frac{
1/4
}
{
( \sqrt{N+1} - \sqrt{N} )^2 + \left( \frac{1 - i\omega\tau}{\sqrt{2}\Omega_B\tau} \right)^2 
}
\label{s0_xy}\\
&&
+
\frac{
1/4
}
{
( \sqrt{N+1} + \sqrt{N} )^2 + \left( \frac{1 - i\omega\tau}{\sqrt{2}\Omega_B\tau} \right)^2 
}
\Biggr],
\nonumber
\end{eqnarray}
\begin{eqnarray}
&&
\sigma_{xx}(\omega) = \frac{ e^2  }{ h }\frac{ 1-i\omega\tau }{ \sqrt{2}\Omega_B\tau }\Biggl\{
\label{s0_xx}\\
&&
\times
\frac{
1/4
}
{
(\sqrt{N+1} - \sqrt{N})\biggl[( \sqrt{N+1} - \sqrt{N} )^2 + \left( \frac{1 - i\omega\tau}{\sqrt{2}\Omega_B\tau} \right)^2\biggr] 
}
\nonumber\\
&&
+
\frac{
1/4
}
{
(\sqrt{N+1} + \sqrt{N})\biggl[( \sqrt{N+1} + \sqrt{N} )^2 + \left( \frac{1 - i\omega\tau}{\sqrt{2}\Omega_B\tau} \right)^2\biggr] 
}
\nonumber\\
&&
+
\sum\limits_{n\geq N+1}
\frac{
1/2
}
{
\sqrt{n+1} + \sqrt{n}
}
\times
\nonumber\\
&&
\times 
\frac{
1
}
{
( \sqrt{n+1} + \sqrt{n} )^2 + \left( \frac{1 - i\omega\tau}{\sqrt{2}\Omega_B\tau} \right)^2
}
\Biggr\}.
\nonumber
\end{eqnarray}
In Eq. (\ref{s0_xx}) the first and second terms arise from the intraband ($N \to N+1$) and interband ($-N\to N+1$) transitions 
(see also Fig. \ref{LLs_sym}), while the third term accounts for the interband transitions from the hole LLs $...,-N-2, -N-1$ to the particle LLs 
$N+1, N+2, ...$. The latter transitions contribute to $\sigma_{xx}$ regardless of the presence (or absence) of the particle-hole symmetry.   

For weak magnetic fields when
\begin{equation}
N \approx \frac{h n_s}{e B} \gg 1,
 \label{Classic}
\end{equation}
the main contribution to Eqs. (\ref{s0_xy}) and (\ref{s0_xx}) comes from the intraband $N \to N+1$ transition, leading to the classical 
AC conductivities 
\begin{equation}
\sigma_{xy} \approx \sigma_s  
\frac{\Omega_c\tau }{(\Omega_c\tau)^2 + (1-i\omega\tau)^2 },
\label{s_xy_cl}
\end{equation}
\begin{equation}
\sigma_{xx} \approx \sigma_s
\frac{ 1-i\omega\tau }{(\Omega_c\tau)^2 + (1-i\omega\tau)^2},
\label{s_xx_cl}
\end{equation}
where $\Omega_c$ is the cyclotron frequency corresponding to a resonant transition between the highest occupied ($N$) and lowest unoccupied ($N+1$) LLs 
(see also Fig. \ref{LLs_sym}):
\begin{eqnarray}
\Omega_c =v \sqrt{ \frac{2eB}{\hbar} } ( \sqrt{N+1} - \sqrt{N} ) \approx \frac{eB v}{\hbar k_F} = \frac{eBv^2}{E_F}.
 \label{Omega_c}
\end{eqnarray}
Unlike the quadratic-dispersion case, $\Omega_c$ depends on the surface carrier density, $\Omega_c \propto 1/\sqrt{n_s}$, through the Fermi wave-vector 
$k_F = \sqrt{4\pi n_s}$, while $v$ is the $n_s$-independent band structure parameter.

Complemented by the Maxwell equations, the magnetotransport theory discussed in this section lays the basis for magneto-optical spectroscopy of TIs, 
which is currently in the focus of both experimental \cite{LaForge10,Sushkov10,Shuvaev11,Hancock11,Aguilar12,Shuvaev12} and theoretical 
\cite{Qi08,Essin09,Maciejko10b,Tse10a,Tse10b,GT11_TI,Tse11,Pesin12} research. 
In particular, the Faraday and Kerr effects have been observed in the TI materials \cite{Sushkov10,Shuvaev11,Hancock11,Aguilar12,Shuvaev12}. 
In high-mobility strained 3DTI HgTe, the Faraday effect reveals the QHE oscillations with low LL indices $n \sim 1$ \cite{Shuvaev12}. 
The broken time-reversal symmetry of the surface QH states has been predicted to give rise to rich magnetoelectric phenomena 
specific to axion electrodynamics \cite{Qi08,Essin09,Maciejko10a,Tse10a,GT11_TI}.
\begin{figure}[t]%
\begin{center}  
\includegraphics[width=0.8\linewidth]{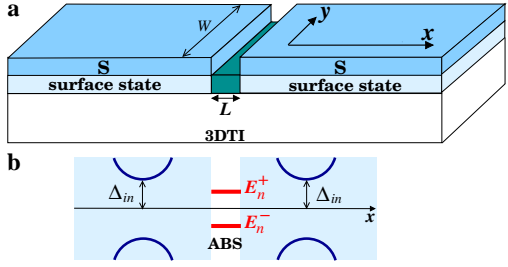}%
\end{center}
\caption{%
(a) S/TI/S Josephson junction and (b) schematic energy diagram of the junction, 
showing Andreev bound states (ABSs) within induced superconducting gap $\Delta_{in}$ on TI surface (see also text). 
}
\label{STIS}
\end{figure}

\section{Superconducting Klein tunneling on topological surfaces.}
\label{Super}

Recently, experiments on the TIs have advanced to the investigation of superconducting junctions involving 3DTIs as a weak link between conventional 
(singlet s-wave) superconductors (Ss) \cite{D_Zhang11,Koren11,Sacepe11,Veldhorst12,Wang12_STI,Williams12} (see also Fig. \ref{STIS}). 
The novelty of such S/TI/S junctions is intimately related to the electron spin helicity: 
since for the helical states the spin-rotation symmetry is broken, 
a conventional singlet s-wave S (e.g. Nb or Al) is expected to induce not only the singlet s-wave pairing, 
but also the unconventional triplet p-wave pairing on the surface underneath the superconductor 
\cite{Fu08,Santos10,Stanescu10,Potter11,Labadidi11,Tanaka12}.

The key question that arises at this stage is what are the potentially observable phenomena that could serve as 
a “smoking gun” for the p-wave superconductivity in S/TI/S systems? One of such phenomena, widely discussed in literature, 
could be the formation of Majorana bound states, topological midgap states that are known to appear on edges (or in a vertex core) 
of a p-wave S (see e.g. Refs. \cite{Read00,Ivanov01,Kitaev01,Nayak08,Oreg10,Qi11,Beenakker11,Tanaka12,Mourik12}). 
The Majorana bound states are expected to give rise to unconventional Josephson effects \cite{Kitaev01,Kwon04,Ioselevich11,Jiang11}, 
dc superconducting tunneling and current noise \cite{Badiane11}. 
For the detection of the Majorana bound states it is essential, however, to avoid hybridization between different Majorana states. 
In Josephson junctions this requirement is not easy to fulfill because when two Ss are brought in contact 
their midgap edge states would normally hybridize to become a pair of Andreev bound states (ABSs) with finite energies 
depending on the phase difference between the Ss (see e.g. Ref. \cite{Kwon04} for the case of intrinsic p- and d-wave Ss). 
In such a situation another closely related phenomenon - the superconducting Klein tunneling - 
can serve as a “smoking gun” for the p-wave superconductivity in S/TI/S junctions. Below we discuss this in more details.

\subsection{ Generalized Fu-Kane model of superconducting proximity effect in 3DTIs.}
\label{Pairing}
We begin by reviewing the superconducting proximity effect for a single lateral contact between a conventional singlet s-wave S 
and the surface of a 3DTI (see also Fig. \ref{STIS}). Such a hybrid system was first considered by Fu and Kane \cite{Fu08}. 
In their approach the proximity effect on the surface is described by a singlet pairing potential treated phenomenologically 
as an energy-independent constant. On the other hand, microscopic approaches (e.g. McMillan's model \cite{McMillan68}) 
allow for a more general energy-dependent description of the proximity effect in terms of the Green's functions of the S. 
We will therefore follow McMillan's model \cite{McMillan68} and its adaptations to low-dimensional systems 
(see e.g. \cite{Grajcar02,GT04,Fagas05,Kopnin11}). In this model, the coupling between the systems is described by a tunneling Hamiltonian, 
allowing one to calculate the Green's function of the normal system, $\hat{G}_{\bf k}$, by summing up relevant Feynman diagrams generated 
by the tunneling Hamiltonian. The superconducting proximity is accounted for by a tunneling self-energy $\hat{\Sigma}(\epsilon)$
in the equation for $\hat{G}_{\bf k}$: 
\begin{equation} 
[\epsilon - \hat{H}_{\bf k} - \hat{\Sigma}(\epsilon) ]\hat{G}_{\bf k}= \hat{I}, \quad 
\hat{H}_{\bf k}=
\Biggl[
\begin{array}{cc}
h_{\bf k} &  0  \\
0 &  -h^*_{-\bf k}
\end{array}
\Biggr],
\label{Eq_G}
\end{equation}
where $\hat{H}_{\bf k}$ is the Hamiltonian of the surface state in $2 \times 2$ Nambu (particle-hole) representation, 
with $h_{\bf k} = \hbar v \mbox{\boldmath$\sigma$}\cdot {\bf k} - E_F$ being itself a $2 \times 2$ matrix in spin space. 
The self-energy $\hat{\Sigma}(\epsilon)$ is also a matrix in Nambu space with the following structure \cite{McMillan68}: 
\begin{equation}
\hat{\Sigma}(\epsilon) =
\left[
\begin{array}{cc}
-i\Gamma(\epsilon)\sigma_0   &  \Delta_{_N}(\epsilon) i\sigma_y {\rm e}^{i\chi}  \\
-\Delta_{_N}(\epsilon) i\sigma_y {\rm e}^{-i\chi}   &  -i\Gamma(\epsilon) \sigma_0  
\end{array}
\right].
\label{Sigma}
\end{equation}
Its off-diagonal elements yield the induced singlet pairing potential 
($\chi$ is the phase of the superconducting order parameter), 
while the diagonal elements in Eq. (\ref{Sigma}) account for the spectrum shift due to the tunneling: 
\begin{eqnarray}
&&
\Delta_{_N}(\epsilon) = i \Gamma_{_N} f_{_S}( \epsilon ) =  i \Gamma_{_N} \frac{ \Delta_{_S} }{ \sqrt{ \epsilon^2 - \Delta^2_{_S} } }, 
\label{Delta_N}\\
&&
\Gamma(\epsilon) = \Gamma_{_N} g_{_S}( \epsilon ) = \Gamma_{_N} \frac{ \epsilon }{ \sqrt{ \epsilon^2 - \Delta^2_{_S} } }, 
\,\, 
\Gamma_{_N} =\pi t^2 N_{_S}.
\label{Gamma_N}
\end{eqnarray}
Here the tunneling energy scale is given by $\Gamma_{_N}$ [it determines the normal-state escape rate into the S, 
$t$ is the tunneling coupling strength, and $N_{_S}$ is the normal-state density of states at the Fermi level in S]. 
$f_{_S}( \epsilon )$ and $g_{_S}( \epsilon )$ are, respectively, the condensate and single-particle quasi-classical Green's functions 
of the S [$\Delta_{_S}$ is the gap energy in S]. From Eqs. (\ref{Eq_G}) and (\ref{Sigma}) we explicitly find the effective 
surface-state Hamiltonian with induced pairing:  
\begin{eqnarray}
&
\hat{H}^{eff}_{\bf k} = \hat{H}_{\bf k} + \hat{\Sigma}(\epsilon) = 
&
\label{H_eff}\\
& 
\left[
\begin{array}{cc}
\hbar v \mbox{\boldmath$\sigma$}\cdot {\bf k} - E_F -i\Gamma(\epsilon)   &  \Delta_{_N}(\epsilon) i\sigma_y {\rm e}^{i\chi}  \\
-\Delta_{_N}(\epsilon) i\sigma_y {\rm e}^{-i\chi}   &  \hbar v \mbox{\boldmath$\sigma^*$}\cdot {\bf k} + E_F -i\Gamma(\epsilon)   
\end{array}
\right].
&
\nonumber
\end{eqnarray}
Consequently, one can calculate all the Nambu matrix elements of the surface-state Green's function:
\begin{equation}
\hat{G}_{\bf k} =
\left[
\begin{array}{cc}
G_{ 11|{\bf k} }  &  G_{ 12|{\bf k} }  \\
G_{ 21|{\bf k} }  &  G_{ 22|{\bf k} }
\end{array}
\right],
\label{G}
\end{equation}
where $G_{ 11(22)|{\bf k} }$ and $G_{ 12(21)|{\bf k} }$ are the particle (hole) and condensate 
Green's functions, respectively. We will briefly discuss the spin structure of 
the condensate function $G_{ 21|{\bf k} }$ as it reveals induced mixed s + p-wave superconducting 
correlations:
\begin{eqnarray}
&&
G_{ 21|{\bf k} } = \frac{1}{i\hbar} 
\left[
\begin{array}{cc}
\langle a^\dag_{\uparrow {\bf -k} } a^\dag_{\uparrow {\bf k} } \rangle   &  
\langle a^\dag_{\uparrow {\bf -k} } a^\dag_{\downarrow {\bf k} } \rangle   \\
\langle a^\dag_{\downarrow {\bf -k} } a^\dag_{\uparrow {\bf k} }  \rangle &  
\langle a^\dag_{\downarrow {\bf -k} } a^\dag_{\downarrow {\bf k} } \rangle 
\end{array}
\right]
=
\label{Spin_pairing}\\
&&
=-\frac{  \frac{1}{2}(\sigma_0 + \mbox{\boldmath$\sigma$}\cdot \hat{\bf k} )  \, i\sigma_y \Delta_{_N}\, {\rm e}^{-i\chi} } 
                 { (\epsilon + i\Gamma(\epsilon) )^2 - \hbar^2 v^2(k-k_F)^2 - 
                 \Delta^2_{_N}(\epsilon) }.
\label{G_21}
\end{eqnarray}
Here we compare the general spin structure of $G_{ 21|{\bf k} }$ (\ref{Spin_pairing}) involving 
the ground-state expectation values of all time-ordered pairs of the creation operators $a^\dagger$
with the explicit solution (\ref{G_21}) obtained from Eq. \ref{Eq_G}. 
Note that in addition to the singlet component $\sim i\sigma_y $, 
the spin-helicity $\mbox{\boldmath$\sigma$}\cdot \hat{\bf k}$ generates the spin-triplet p-wave component 
$\sim \mbox{\boldmath$\sigma$}\cdot \hat{\bf k} i\sigma_y $ with the same strength $\Delta_{_N}$ 
[$\hat{\bf k}$ is the unit vector in the momentum direction on the Fermi surface].      
Additionally, from the denominator of Eq. (\ref{G_21}) we see that an isotropic energy gap $\Delta_{in}$ 
is induced in the surface spectrum (see also Fig. \ref{STIS}):
\begin{equation}
\Delta_{in} \approx |\Delta_{_N}(0)| = \Gamma_{_N}, \qquad \epsilon, \Gamma_{_N} \ll \Delta_{_S}.
\label{Gap}
\end{equation}
The origin of the mixed s- and p-wave superconducting correlations in Eq. (\ref{G_21}) is the broken spin-rotation symmetry 
of the helical surface state. The situation reminds the mixed singlet -triplet intrinsic superconductivity predicted in Ref. \cite{Gorkov01} 
for systems without inversion symmetry. Unlike intrinsic p-wave superconductors, 
the conservation of the spin helicity in TIs guaranties the robustness the s + p - wave proximity effect against potential impurity scattering 
\cite{Potter11}. We also note that Eq. (\ref{G_21}) retains the mixed s + p - wave structure in the limit $\epsilon/\Delta_{_S} \to 0$, 
which corresponds to the case of Ref. \cite{Fu08}.

\subsection{Superconducting Klein tunneling and topological Andreev bound states.}
\label{Klein_ABS}

We consider now a short weak link with length $L \ll  \hbar v/\Gamma_{_N}$, choosing the superconductor phases on the left and right as 
$\chi_L =0$ and $\chi_R = \chi$. At the simplest level the junction can be described by the following real-space equations:
\begin{eqnarray}
&&
[\hat{H}^{eff}( {\bf r} ) + \tau_z\sigma_0 U \delta(x) ] \Psi({\bf r}) = \epsilon \,\Psi({\bf r}), 
\label{Eq_Psi}\\
&&
\Psi(x,y=0) = \Psi(x,y=W),
\label{PBC}
\end{eqnarray}
for the Nambu spinor 
$\Psi = [(\Psi_{1\uparrow},\Psi_{1\downarrow}), (\Psi_{2\uparrow}, \Psi_{2\downarrow})]^T$ 
which combines the particle $(\Psi_{1\uparrow},\Psi_{1\downarrow})^T$ 
and hole $(\Psi_{2\uparrow}, \Psi_{2\downarrow})^T$ spinors. 
The transverse periodic boundary conditions (\ref{PBC}) can be realized 
in a surface array of Josephson junctions. In order to model the interface scattering inside the weak link we introduce the potential 
$U \delta(x)$ (see Ref. \cite{Olund12} for the scattering-free case). The solution of Eq. (\ref{Eq_Psi}) can be sought as the sum of independent 
channels $\Psi({\bf r}) = \sum_{k_n} \Psi_{k_n}(x) {\rm e}^{ik_n y}/\sqrt{W}$, with $k_n=2\pi n/W$ and $n \in Z$. 
Searching for ABSs with $\Psi_{k_n}(x\to \pm \infty) \to 0$ we obtain the eingevalue equation:
\begin{equation}
\left(  \frac{\epsilon}{ \Delta_{_S} } + \frac{\epsilon}{ \Gamma_{_N} } \sqrt{  1 - \frac{\epsilon^2}{ \Delta^2_{_S} }  } \right)^2 
= 1 - T_n\sin^2 \frac{\chi}{2}, 
\label{Eq_E}
\end{equation}
where $T_n$ is the $n$'s channel normal-state transparency:  
\begin{equation}
T_n = \frac{ 1 - (k_n/k_F)^2 } 
           { 1 - (k_n/k_F)^2/(1+u^2) }, \quad u = \frac{U}{\hbar v},
\label{T}    
\end{equation}
with $|k_n|\leq k_F$, determining the number of the open channels in the junction, $n_{ch}$. 
Eq. (\ref{Eq_E}) accounts for the energy dependence of both $\Delta_{_N}(\epsilon)$ and $\Gamma(\epsilon)$.

\begin{figure}[t]%
\begin{center}  
\includegraphics[width=0.85\linewidth]{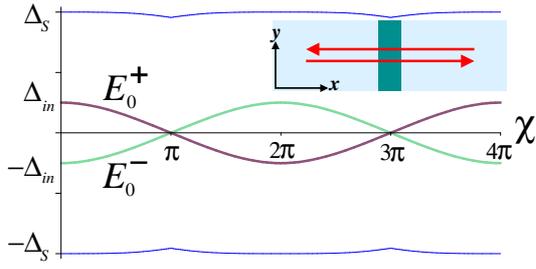}%
\end{center}
\caption{%
Topological gapless ABSs $E^\pm_0(\chi)$ (\ref{E_0}). They propagate perpendicular to junction barrier without backscattering. 
These states are orthogonal and $4\pi$ periodic (see also text). 
}
\label{ABS_0}
\end{figure}

Superconducting Klein tunneling and topological ABSs occur in the $n=0$ channel, 
as it propagates perpendicularly to the junction barrier and is protected against backscattering ($T_0=1$, see also Fig. \ref{ABS_0}). 
We obtain the ABS spectrum from Eq. (\ref{Eq_E}) using expansion in small parameter 
$\gamma = \Gamma_{_N}/\Delta_{_S}\ll 1$: 
\begin{eqnarray}
E^\pm_{n=0}(\chi) \approx \pm 
\Gamma_{_N} \left[  (  1 - \gamma + \gamma^2  ) \cos\frac{\chi}{2} + 
\frac{\gamma^2}{2} \cos^3\frac{\chi}{2} \right]. \,\,
\label{E_0}
\end{eqnarray}
These states have orthogonal wave functions. Their spinor structure is simplest at zero energy: 
\begin{eqnarray}
&&
\Psi_+(x)\bigl|_{E^+_0 \to 0} = 
\left( 
{\tiny
\left[
\begin{array}{c}
 1 \\ 1\\ 0\\ 0
\end{array}
\right]
-i 
\left[
\begin{array}{c}
 0 \\ 0\\ 1\\ -1
\end{array}
\right]
}
\right) \frac{  {\rm e}^{ ik_Fx - x/\xi }  }{ 2\sqrt{\xi} }, 
\label{Psi_+}\\
%
%
&&
\Psi_-(x)\bigl|_{E^-_0 \to 0} = 
\left( 
{\tiny
\left[
\begin{array}{c}
 1 \\ -1\\ 0\\ 0
\end{array}
\right]
-i 
\left[
\begin{array}{c}
 0 \\ 0\\ 1\\ 1
\end{array}
\right]
}
\right) \frac{  {\rm e}^{ -ik_Fx - x/\xi }  }{ 2\sqrt{\xi} }, 
\label{Psi_-}
\end{eqnarray}
where $\xi = \hbar v/\Gamma_{_N}$ is the healing length.
The orthogonality of (\ref{Psi_+}) and (\ref{Psi_-}) prevents mixing of the counter-propagating modes, yielding the gapless $4\pi$-periodic ABS spectrum (\ref{E_0}). 
Moreover, the particle (\ref{Psi_+}) and hole (\ref{Psi_-}) midgap states are constructed of the same two spinors $[1,1]^T$ and $[1,-1]^T$ 
describing two degenerate orthogonal Majorana modes. Similar midgap states were found in the core of a vertex on the 3DTI surface \cite{Fu08} 
(see also reviews \cite{Hasan10,Qi11,Tanaka12,Beenakker11} for more details). 

\begin{figure}[t]%
\begin{center}  
\includegraphics[width=0.85\linewidth]{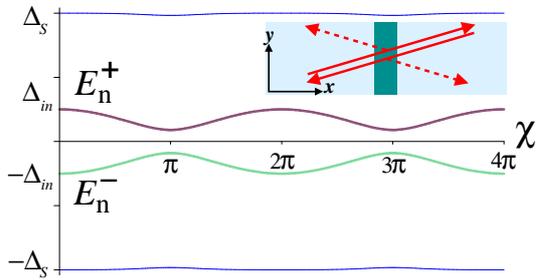}%
\end{center}
\caption{%
Nontopological ABSs $E^\pm_n(\chi)$ (\ref{E_n}) correspond to oblique incidence at junction barrier. 
Scattering from barrier results in gapped $2\pi$-periodic spectrum (see also text). 
}
\label{ABS_n}
\end{figure}

We should emphasize the difference between the topological ABSs (\ref{E_0}) and those appearing 
in p-wave Josephson junctions (see e.g. Ref. \cite{Kwon04}). 
First, because of the superconducting Klein tunneling 
the ABSs (\ref{E_0}) are completely independent of the details of the junction barrier. 
Second, the energy dependence of the proximity effect leads to the cubic and, in general, 
to higher odd-power terms in the spectrum. Additionally, due to the energy dependence of the proximity effect Eq. (\ref{Eq_E}) 
has high-energy solutions appearing just below the superconductor gap $\Delta_{_S}$
(see thin blue curves in Figs. \ref{ABS_0} and \ref{ABS_n}; these solutions will be discussed in detail elsewhere).

The transport channels with $k_n\not =0$ support nontopological ABSs because, 
at oblique incidence, scattering from the junction barrier generates an energy gap in the spectrum, 
making it $2\pi$-periodic (see also Fig. \ref{ABS_n}):
\begin{eqnarray}
E^\pm_{n}(\chi) \approx &\pm& \Gamma_{_N} 
\left[  (  1 - \gamma + \gamma^2  ) \sqrt{ 1 - T_n \sin^2(\chi/2) } + \right.
\nonumber\\
& +&
\left.
(\gamma^2/2) \left[1 - T_n \sin^2(\chi/2) \right]^{3/2} \right].
\label{E_n}
\end{eqnarray}
The topological (\ref{E_0}) and usual (\ref{E_n}) ABSs give rise to a peculiar AC Josephson current discussed below.

\subsection{Fractional AC Josephson effect.}
\label{AC_Jo}

The AC Josephson effect is observed in voltage-biased junctions where the supercurrent oscillates 
with the frequency $\omega_{_V}=2eV/\hbar$ proportional to the bias voltage $V$ and Cooper pair charge $2e$ \cite{Yanson65,Langenberg65}. 
In weak links between intrinsic p-wave superconductors, 
the $4\pi$-periodic ABSs are expected to give rise to an unconventional AC Josephson effect at the fractional frequency $\omega_{_V}/2$ 
(see e.g. Refs. \cite{Kitaev01,Kwon04}). Leaving aside calculations (see Ref. \cite{Kwon04} for details) 
we present the result for the AC Josephson current in our proximity S/TI/S system:
\begin{equation}
J(t) \approx  J_0\left( \omega_{_V} t  \right)   + 
              J_1 \sin \frac{ \omega_{_V} }{2} t + J_2 \sin \frac{ 3\omega_{_V} }{2} t +..., 
\label{J_t}
\end{equation}
Here the function $J_0\left( \omega_{_V} t  \right)$ is the conventional AC Josephson current carried by the $2\pi$-periodic ABSs (\ref{E_n}), 
while the other two terms arise from the topological $4\pi$-periodic ABSs (\ref{E_0}). 
Unlike the p-wave junctions \cite{Kitaev01,Kwon04}, not only the main fractional frequency $\omega_{_V}/2$, but also $3\omega_{_V}/2$ and, in general, 
higher odd harmonics appear due to the energy dependence of the proximity-induced $\Delta_{_N}(\epsilon)$ (\ref{Delta_N}) and $\Gamma(\epsilon)$ (\ref{Gamma_N}). 
The estimates of $J_0, J_1$ and $J_2$ are given by
\begin{eqnarray}
&&
J_0 \sim  \frac{n_{ch} e \Gamma_{_N}}{\hbar} (1- \gamma + \gamma^2), 
\label{J0}\\
&& 
J_1 \sim \frac{e \Gamma_{_N} }{\hbar} \left(1- \gamma + \frac{11}{8}\gamma^2 \right), \quad 
J_2 \sim \frac{e \Gamma_{_N} }{\hbar} \frac{ 3\gamma^2 }{8}.
\label{Js}
\end{eqnarray}
We can conclude that favourable conditions for the observation of the topological ABSs 
exist in narrow weak links with a small number of open channels $n_{ch} \sim 1$.

\begin{acknowledgement}
We thank L. W. Molenkamp, S.-C. Zhang, A. H. MacDonald, H. Buhmann, C. Br\"une, J. Oostinga, B. Trauzettel, P. Recher, E. G. Novik, P. Virtanen, 
A. Pimenov, G. V. Astakhov, K. Richter, P. M. Ostrovsky, A. D. Mirlin, C.-X. Liu, M. Guigou, P. Michetti and J. Budich for many valuable discussions.  
This work was financially supported by the German research foundation DFG [Grant No. HA5893/3-1].
\end{acknowledgement}

%
%

\end{document}